\documentclass[7pt,twocolumn]{article}
\usepackage[margin=0.53in]{geometry}
\usepackage{tgpagella}
\usepackage[T1]{fontenc}
\usepackage{tgcursor}

\date{}
\usepackage{caption}
\usepackage{amsmath}
\usepackage{amsfonts}
\usepackage[numbers,sort&compress]{natbib}

\makeatletter
\renewcommand\section{%
  \@startsection{section}{1}{\z@}%
    {2ex plus .2ex minus .1ex}
    {1ex plus .2ex}
    {\normalfont\large\bfseries}  
}%
\makeatother
\makeatletter
\renewcommand\subsection{%
  \@startsection{subsection}{1}{\z@}%
    {2ex plus .2ex minus .1ex}
    {1ex plus .2ex}
    {\normalfont\bfseries}  
}%
\makeatother

\usepackage[caption=false]{subfig} 
\usepackage{booktabs}
\usepackage{colortbl}
\usepackage{multirow}
\usepackage{enumitem}
\usepackage{siunitx}
\usepackage{musicography}
\usepackage{fancyvrb}
\usepackage{listings}
\usepackage{makecell}

\usepackage{xcolor}
\definecolor{codegreen}{rgb}{0,0.6,0}
\definecolor{codegray}{rgb}{0.5,0.5,0.5}
\definecolor{codepurple}{rgb}{0.58,0,0.82}
\definecolor{backcolour}{rgb}{0.96,0.96,0.92}

\usepackage{bm}
\usepackage{hyperref}

\usepackage{algorithm}
\usepackage{algpseudocode}
\algrenewcommand\algorithmicrequire{\textbf{Input:}}
\algrenewcommand\algorithmicensure{\textbf{Output:}}
\algnewcommand\algorithmicinput{\textbf{Input:}}
\algnewcommand\algorithmicoutput{\textbf{Output:}}
\algnewcommand\Input{\item[\algorithmicinput]}%
\algnewcommand\Output{\item[\algorithmicoutput]}%

\DeclareMathAlphabet{\mathcal}{OMS}{cmsy}{m}{n}

\renewcommand\paragraph[1]{\noindent\textbf{#1 ---}}
\newcommand\Tc{T_\mathrm{pool}}
\newcommand\bfT{\mathbf{T}}
\newcommand\bbfT{\bar{\mathbf{T}}}

\newcommand\bfIG{\mathbf{I}^\mathrm{G}}
\newcommand\bfIA{\mathbf{I}^\mathrm{A}}
\newcommand\bfIP{\mathbf{I}^\mathrm{P}}
\newcommand\bfIS{\mathbf{I}^\mathrm{S}}
\newcommand\bfE{\mathbf{E}}
\newcommand\bfU{\mathbf{U}}
\newcommand\bbfU{\bar{\mathbf{U}}}
\newcommand\bfS{\mathbf{S}}

\newcommand\bfP{\mathbf{P}}
\newcommand\bfY{\mathbf{Y}}

\newcommand\Gc{G_\mathrm{c}}
\newcommand\Gp{G_\mathrm{p}}
\newcommand\Vc{V_\mathrm{cand}}

\newcommand\Vp{V_\mathrm{p}}
\newcommand\bfw{\mathbf{w}}
\newcommand\bfm{\mathbf{m}}

\newcommand\La{L_\mathrm{a}}
\newcommand\Lr{L_\mathrm{r}}
\newcommand\Lg{L_\mathrm{g}}
\newcommand\Lp{L_\mathrm{p}}
\newcommand\fn[1]{\mathrm{#1}\:\!}
\newcommand\textttt[1]{\text{\scriptsize$\texttt{#1}$}}

\DeclareMathOperator*{\argmin}{arg\,min}

\usepackage{tikz}
\usetikzlibrary{external}
\usetikzlibrary{arrows.meta, bending, decorations.pathmorphing}
\definecolor{emerald}{rgb}{0.31, 0.78, 0.47}

\begin{document}


\title{Reverse Engineering of Music Mixing Graphs with \\ Differentiable Processors and Iterative Pruning}

\author{
Sungho Lee$^1$, Marco A. Martínez-Ramírez$^2$, Wei-Hsiang Liao$^2$, Stefan Uhlich$^3$, \\ 
Giorgio Fabbro$^3$, Kyogu Lee$^1$, and Yuki Mitsufuji$^{2,4}$ \\
}
\date{{\small $^1$Department of Intelligence and Information, Seoul National University, Seoul, South Korea \\
\vspace{-.5mm}
$^2$Sony AI, Tokyo, Japan \quad $^3$Sony Europe B.V., Stuttgart, Germany \quad $^4$Sony Group Corporation, Tokyo, Japan}
}

\twocolumn[
  \begin{@twocolumnfalse}
    \maketitle
    \noindent
    \begin{center}
    \begin{minipage}{0.77\textwidth}
    \textbf{Abstract ---}
    Reverse engineering of music mixes aims to uncover how dry source signals are processed and combined to produce a final mix. 
    We extend the prior works to reflect the compositional nature of mixing and search for a graph of audio processors. 
    First, we construct a mixing console, applying all available processors to every track and subgroup.
    With differentiable processor implementations, we optimize their parameters with gradient descent. 
    Then, we repeat the process of removing negligible processors and fine-tuning the remaining ones.
    This way, the quality of the full mixing console can be preserved while removing approximately two-thirds of the processors.
    The proposed method can be used not only to analyze individual music mixes but also to collect large-scale graph data that can be used for downstream tasks, e.g., automatic mixing.
    Especially for the latter purpose, efficient implementation of the search is crucial. To this end, we present an efficient batch-processing method that computes multiple processors in parallel. We also exploit the “dry/wet” parameter of the processors to accelerate the search. 
    Extensive quantitative and qualitative analyses are conducted to evaluate the proposed method's performance, behavior, and computational cost.
    \end{minipage}
    \end{center}
    \bigskip
    \vspace{5mm}
  \end{@twocolumnfalse}%
]

\section{Introduction}
Production of modern music recordings involves a mixing process, balancing source tracks in various aspects such as loudness, frequency content, and spatialization.
Mixing engineers achieve this by combining and controlling audio processors, such as equalizers and artificial reverberation.
Numerous attempts have been made to uncover this complex process. One way is by directly collecting expert knowledge \cite{pestana2014intelligent, everardo2017towards, ronan2017analysis}.
Alternatively, one can develop a computational method that estimates the mixing process from input source tracks and output mix. 
This strategy, so-called \emph{reverse engineering}, has several advantages. 
First, it is useful on its own as it allows one to analyze and understand the individual mixes.
Furthermore, we can collect large-scale mixing data from various public music datasets \cite{bittner2014medleydb, bittner2016medleydb, senior2018mixing} that provide mixes along with their source tracks.
This enables more scalable, data-driven downstream applications. 
For example, we can conduct statistical analysis of the mixing practices \cite{wilson2015101, mourgela2024exploring, wilson2016variation}.
The data can also be used to develop automatic mixing systems \cite{de2017ten, perez2009automatic, de2013knowledge, steinmetz2020diffmixconsole, lee2023blind}, which could include neural network components.

Reverse engineering of music mixing was initially tackled by Barchiesi and Reiss \cite{barchiesi2010reverse}. By assuming the mixing as a linear process, least squares solutions, i.e., an impulse response and a gain envelope per track, were derived and further converted to the processor parameters.
More recently, Colonel and Reiss \cite{colonel2021reverse, colone2023reverse} employed differentiable signal processing \cite{engel2020ddsp, hayes2023review} and directly optimized the processor parameters with gradient descent.
This approach allows more flexible choices of processors and optimization objectives, which could lead to better perceptual matches.

Ideally, the reverse engineering method should be able to comprehensively recover the real-world practice of music mixing.
However, these prior works assumed a fixed chain of processors per source track and estimated only their parameters.
Therefore, it is desirable to extend these methods to also estimate the compositional aspect of mixing, which can be represented with a \emph{graph} of audio processors.

\input{fig-framework-full}

\subsection{Problem Formulation}
We describe the mixing process with a graph $G=(V, E)$ where $V$ and $E$ denote the node and edge set, respectively.
Each node $v_i \in V$ represents either a processor or an auxiliary module, distinguished by a type attribute $t_i$. 
Each edge $e_{ij} \in E$ is a ``cable'' 
that sends an output signal to another node as input. 
We restrict the graph $G$ to be acyclic; hence, it is a directed acyclic graph (DAG).

We consider $7$ processor types: equalizer \texttt{e}, compressor \texttt{c}, noisegate \texttt{n}, stereo imager \texttt{s}, gain/panning \texttt{g}, multitap delay \texttt{d}, and reverb \texttt{r}. 
Each processor takes an input audio $u_i$ and a parameter vector $p_i$ and computes an output $y_i$. 
The auxiliary modules include an input \texttt{i} that outputs a dry source $s_k$, a mix \texttt{m} that sums the incoming signals, and an output \texttt{o} whose sum of the inputs is the final output $\hat{y}$.
We denote a series of nodes with a string, e.g., \texttt{ecns}.

The reverse engineering task is defined as follows: 
Given the dry sources $\bfS = [s_1, \cdots, s_K]$ and the ground-truth mix $y$, our objective is to identify a graph $G$ along with all processor parameters $\bfP$ such that processing the sources produces a mix $\hat{y} = G(\mathbf{S}; \mathbf{P})$ that closely approximates the original mix $y$. To measure the similarity, we employ an audio-domain loss function $\La$. 
In addition, we need some regularization $\Lr$ to restrict the graph and parameters to be in a reasonable range (e.g., not too large).
Therefore, the full optimization objective is given as 
\begin{equation} \label{eq:objective}
        G^*, \bfP^* = 
        \argmin_{G, \bfP}
        \big[ L_\mathrm{a} (\hat{y}, y) + L_\mathrm{r} (G, \bfP) \big].
\end{equation}

\subsection{Proposed Strategy}
Exploring candidate graphs without restrictions might seem appealing.
However, the combinatorial nature of graphs results in a vast search space, and the need to simultaneously optimize processor parameters further complicates the task. 
Also, many graph-parameter pairs can achieve similar match quality, making the task ill-posed and underdetermined. 
Therefore, it is beneficial to impose restrictions, e.g., favoring commonly used structures, to narrow the search space and improve practicality.

To this end, we propose a search method based on pruning (See Figure \ref{fig:framework-full}).
Following Colonel and Reiss \cite{colonel2021reverse, colone2023reverse}, we first create a ``mixing console'' $G_\mathrm{c}$.
To each input source, the mixing console applies a processor chain, which consists of all the $7$ processors in the fixed order.
It subgroups the processed sources with mix \texttt{m}, applies the chain again, and sums the processed subgroups to obtain a final mix $\hat{y}$.
We assume that the subgrouping information is known a priori. Note that this structure resembles the traditional hybrid mixing console \cite{hybridconsole}. 
We follow the differentiable signal processing framework \cite{engel2020ddsp, hayes2023review} 
and implement the entire mixing in the automatic differentiation framework, i.e., \texttt{PyTorch} \cite{paszke2019pytorch}.
This allows us to optimize all console parameters $\bfP_\mathrm{c}$ with given objectives via gradient descent. 
After this, we proceed to the pruning stage, where we reduce the number of processors as much as possible while the match quality remains within a predefined tolerance $\tau$.
The pruning of a node is defined as its removal and the rerouting of its edges
in a way that is equivalent to setting them to ``bypass.''
We adopt the iterative method \cite{castellano1997iterative} proposed for neural network pruning. That is, we optimize the remaining parameters after the pruning and repeat this process multiple times.

Casting the graph search into pruning reduces the search space, 
which has several tradeoffs.
First, the pruning only removes the processors from the pre-defined processor chains, i.e., it does not consider all other possibilities. 
As a result, the pruned graph $G_\mathrm{p}$ does not surpass the original mixing console $\Gc$ in match quality. 
Nevertheless, it aligns with real-world practices by selectively applying appropriate processors, promoting sparsity, and enhancing the interpretability of the estimated mixing process.
Also, it significantly reduces the search cost and allows us to obtain a practical algorithm, which might be challenging with other alternatives \cite{liu2018darts, ye2023fm}.
Our evaluation results show that about $67\%$ of the processors can be removed while having little perceptual degradation.

Making the search algorithm fast and efficient is crucial to applying it to large-scale mixing data. 
To achieve this, we employ two strategies. 
First, our search repeatedly computes the mix, which is a major source of bottleneck. 
Hence, we derive an algorithm that accelerates this process by computing multiple processors within a graph in parallel. 
This differs from prior implementations \cite{engel2020ddsp, uzrad2024diffmoog} that compute every processor ``one by one." Note that our graphs are much larger and more parallelizable than those from previous works \cite{engel2020ddsp, uzrad2024diffmoog, ye2023fm, lee2023blind}, yielding significant speedup with our method.
Second, pruning is another source of bottleneck because it includes many trials, each involving audio processing and loss evaluations. Thus, we focus on developing efficient pruning methods. In this process, we aim to identify a subset of nodes that minimally impact match quality. To this end, we interpret each processor’s “dry/wet” parameter as an approximate importance score and leverage it to select pruning candidates.
We combine this method with a brute-force node-by-node check, deriving a hybrid method that has a reasonable trade-off between the computational cost and resulting sparsity.

The proposed search method was initially presented in a conference paper \cite{lee2024searching}, and some technical details 
were provided in a companion demonstration paper \cite{lee2024grafx}. 
This journal paper combines these two, aiming to provide a complete description and unified viewpoint of the framework. 
Furthermore, we conduct extended evaluations. 
We validate our methods with a diverse set of objective metrics \cite{kilgour2018fr, le2019sdr, vanka2024diff, man2014analysis}, making the quantitative analysis more rigorous.
Additional experiments and case studies also reveal the behavior of the pruning algorithm in more detail.
We also report subjective listening test results \cite{mushra, schoeffler2018webmushra}, which reveal how the mixing consoles and their pruned versions perform in perceptual similarity. 
The results are further analyzed to reveal the relationship between subjective similarity and widely used objective metrics.
These provide a deeper understanding, including the strengths and weaknesses of the proposed method, suggesting potential areas for improvement.

\subsection{Open-Source Implementation} 
We provide two supplementary materials. The first one, relevant to Section \ref{section:processor-details} and \ref{section:diff_graph}, covers the efficient computation of audio processing graphs with \texttt{PyTorch}.
It was initially developed for the presented pruning task and is now fully extended as a general-purpose library\footnote{\href{https://sh-lee97.github.io/grafx}{\texttt{https://sh-lee97.github.io/grafx}}}.
In this paper, we discuss the main technical details used for the pruning.
Samples and implementation of the graph pruning (Section \ref{section:mixing_consoles}, \ref{section:search}, and \ref{section:pruning-results}) are available in the separate material\footnote{\href{https://sh-lee97.github.io/grafx-prune}{\texttt{https://sh-lee97.github.io/grafx-prune}}}.

\section{Related Works}\label{setion:related_works}

\subsection{Composition of Audio Processors}\label{section:related_graph}
\looseness=-1
\begin{table*}
\setlength\tabcolsep{2.8pt}
\renewcommand{\arraystretch}{.85}
\begin{center}
\fontsize{8.3}{8.3}\selectfont
\caption{
A brief summary and comparison of previous works on the estimation of compositional audio signal processing.  
}
\begin{tabular}{llp{15cm}}
\toprule
\arrayrulecolor{lightgray}
\cite{uzrad2024diffmoog}
& \emph{Task \& domain} & Sound matching $[x]\to[\bfP]$. The synthesizer parameters $\bfP$ were estimated to match the reference (target) audio $x$. \\
\cmidrule{2-3}
& \emph{Processors} & Oscillators, envelope generators, and filters that allow parameter modulation as an optional input. \\
\cmidrule{2-3}
& \emph{Graph} & Any pre-defined directed acyclic graph (DAG). For example, a subtractive synthesizer that comprises $2$ oscillators, $1$ amplitude envelope, and $1$ lowpass filter was used in the experiments. \\
\cmidrule{2-3}
& \emph{Method} & Trained a single neural backbone for the reference encoding, followed by multiple prediction heads for the parameters. Optimized with a parameter loss and spectral loss, where the latter is calculated with every intermediate output. \\
\arrayrulecolor{black}
\midrule
\arrayrulecolor{lightgray}

\cite{caspe2022ddx7}
& \emph{Task \& domain} 
& Sound matching $[x]\to[\bfP]$. A frequency-modulation (FM) synthesizer matches recordings of monophonic instruments (violin, flute, and trumpet). 
Estimates parameters of an operator graph that is empirically searched \& selected. \\
\cmidrule{2-3}
& \emph{Processors} 
& Differentiable sinusoidal oscillators, each used as a carrier or modulator, with pre-defined frequencies. 
An additional FIR reverb is added to the FM graph output for post-processing. \\
\cmidrule{2-3}
& \emph{Graph} 
& DAGs with at most $6$ operators. Different graphs for different target instruments.\\
\cmidrule{2-3}
& \emph{Method} 
& Trained a convolutional neural network that estimates envelopes from the target loudness and pitch. \\
\arrayrulecolor{black}
\midrule
\arrayrulecolor{lightgray}

\cite{ye2023fm}
& \emph{Task \& domain} 
& Sound matching $[x]\to [G, \bfP]$. Similar setup to the above \cite{caspe2022ddx7} plus additional estimation of the operator graph $G$. \\
\cmidrule{2-3}
& \emph{Processors} 
& Identical to \cite{caspe2022ddx7}, except for the frequency ratio that can be searched. \\
\cmidrule{2-3}
& \emph{Graph} 
& A subgraph of a supergraph, which resembles a multi-layer perceptron (modulator layers followed by a carrier layer). \\
\cmidrule{2-3}
& \emph{Method} 
& Trained a parameter estimator for the supergraph and found the appropriate subgraph $G$ with an evolutionary search. \\
\arrayrulecolor{black}
\midrule
\arrayrulecolor{lightgray}

\cite{Mitcheltree_2021}
& \emph{Task \& domain} 
& Reverse engineering $[s, y]\to [G, \bfP]$ of an audio effect chain from a subtractive synthesizer (commercial plugin). \\
\cmidrule{2-3}
& \emph{Processors} 
& $5$ audio effects: compressor, distortion, equalizer, phaser, and reverb. Non-differentiable implementations. \\
\cmidrule{2-3}
& \emph{Graph} 
& Chain of audio effects generated with no duplicate types (therefore $32$ possible combinations) and random order. \\ 
\cmidrule{2-3}
& \emph{Method} 
& Trained a next effect predictor and parameter estimator in a supervised (teacher-forcing) manner. \\
\arrayrulecolor{black}
\midrule
\arrayrulecolor{lightgray}

\cite{guo2023automatic}
& \emph{Task \& domain} 
& Blind estimation $[y]\to [G]$ and reverse engineering $[s,y]\to [G]$ of guitar effect chains.  \\
\cmidrule{2-3}
& \emph{Processors} 
& $13$ guitar effects, including non-linear processors, modulation effects, ambiance effects, and equalizer filters. \\
\cmidrule{2-3}
& \emph{Graph} 
& A chain of guitar effects. Maximum $5$ processors and a total of $221$ possible combinations. \\
\cmidrule{2-3}
& \emph{Method} 
& Trained a convolutional neural network with synthetic data to predict the correct combination. \\
\arrayrulecolor{black}
\midrule
\arrayrulecolor{lightgray}

\cite{take2024audio}
& \emph{Task \& domain} 
& Blind estimation $[y]\to [G, \bfP, s]$ of audio effect chains, the original dry signal, and all the intermediate signals. \\
\cmidrule{2-3}
& \emph{Processors} 
& 4 guitar effects: distortion, delay, chorus, and reverb. \\
\cmidrule{2-3}
& \emph{Graph} 
& A chain of audio effects. \\
\cmidrule{2-3}
& \emph{Method} 
& Trained a hybrid waveform-spectrogram transformer that estimates each processing step (type, parameter, and signal). The full inference is done autoregressively. \\
\arrayrulecolor{black}
\midrule
\arrayrulecolor{lightgray}

\cite{steinmetz2020diffmixconsole} 
& \emph{Task \& domain} 
& Automatic mixing $[\bfS]\to[\bfP]$. Estimated parameters of fixed processing chains from source tracks $(K\leq 16)$. \\
\cmidrule{2-3}
& \emph{Processors} 
& $7$ differentiable processors, where $4$ (gain, polarity, fader, and panning) were implemented exactly. 
A combined effect of the remaining $3$ (equalizer, compressor, and reverb) was approximated with a single pre-trained neural network. \\
\cmidrule{2-3}
& \emph{Graph} 
& Tree structure: applied a fixed chain of the $7$ processors for each track, and then summed the chain outputs together. \\
\cmidrule{2-3}
& \emph{Method} 
& Trained a parameter estimator (convolutional neural network) with a spectrogram loss end-to-end. \\
\arrayrulecolor{black}
\midrule
\arrayrulecolor{lightgray}

\cite{ramirez2021differentiable} 
& \emph{Task \& domain} 
& Reverse engineering of music mastering $[s,y]\to[\bfP]$. \\
\cmidrule{2-3}
& \emph{Processors} 
& A multi-band compressor, graphic equalizer, and limiter. Gradient approximated with a finite difference method. \\
\cmidrule{2-3}
& \emph{Graph} 
& A serial chain of the processors. \\
\cmidrule{2-3}
& \emph{Method} 
& Optimized parameters with gradient descent. \\
\arrayrulecolor{black}
\midrule
\arrayrulecolor{lightgray}

\cite{lee2023blind}
& \emph{Task \& domain} 
& Blind estimation $[y]\to[G,\bfP]$ and reverse engineering $[\bfS, y]\to[G, \bfP]$.
Estimates both the graph and its parameters for singing voice effect $(K=1)$ or drum mixing $(K\leq 6)$. \\
\cmidrule{2-3}
& \emph{Processors} 
& A total of $33$ processors, including linear filters, nonlinear filters, and control signal generators. 
Some processors are multiple-input multiple-output (MIMO), e.g., allowing auxiliary modulations. Non-differentiable implementations. \\ 
\cmidrule{2-3}
& \emph{Graph} 
& Complex DAG; splits (e.g., multi-band processing) and merges (e.g., sum and modulation). $30$ processors max. \\
\cmidrule{2-3}
& \emph{Method} 
& Trained a convolutional neural network-based reference encoder and a transformer variant for graph decoding and parameter estimation. 
Both were jointly trained via direct supervision of synthetic graphs (e.g., parameter loss). \\
\arrayrulecolor{black}
\midrule
\arrayrulecolor{lightgray}

\cite{colone2023reverse} 
& \emph{Task \& domain} 
& Reverse engineering $[\bfS, y]\to [\bfP]$ of music mixing. Estimated parameters of a fixed chain for each track. \\
\cmidrule{2-3}
& \emph{Processors} 
& $6$ differentiable processors: gain, equalizer, compressor, distortion, panning, and reverb. \\
\cmidrule{2-3}
& \emph{Graph} & A chain of $5$ processors (all above types except the reverb) for each dry track (any other DAG can also be used). The reverb is used for the mixed sum. \\
\cmidrule{2-3}
& \emph{Method} & Parameters were optimized with spectrogram loss end-to-end via gradient descent. \\
\arrayrulecolor{black}
\midrule
\arrayrulecolor{lightgray}

Ours
& \emph{Task \& domain} 
& Reverse engineering $[\bfS, y]\to [G, \bfP]$ of music mixing. Estimated a chain of processors and their parameters for each track and submix where $K\leq 130$. \\
\cmidrule{2-3}
& \emph{Processors} & $7$ differentiable processors: gain/panning, stereo imager, equalizer, reverb, compressor, noisegate, and delay. \\
\cmidrule{2-3}
& \emph{Graph} & A tree of processing chains with a subgrouping structure. Processors can be omitted, but should follow the fixed order. \\
\cmidrule{2-3}
& \emph{Method} & Joint estimation of the soft masks (dry/wet weights) and processor parameters.
Optimized with the spectrogram loss (and additional regularizations) end-to-end via gradient descent.
Accompanied by hard pruning stages. 
\\
\arrayrulecolor{black}

\bottomrule
\end{tabular}
\label{table:related_works}
\end{center}
\end{table*}

While our work focuses on music mixing, it is valuable to review recent works that incorporate the compositional aspects of musical signal processing in a broader scope, such as sound matching \cite{uzrad2024diffmoog, caspe2022ddx7, ye2023fm}, effect chain estimation \cite{Mitcheltree_2021, guo2023automatic, take2024audio, ramirez2021differentiable, lee2023blind}, and music mixing estimation \cite{steinmetz2020diffmixconsole, lee2023blind, colone2023reverse}. Table \ref{table:related_works} highlights the key differences among these approaches.
These works vary in their tasks, domains, processors, graph structures, and estimation methods.
We denote each task with $[A] \to [B]$ where $A$ and $B$ denote given references and estimation targets, respectively.
For example, if the references are dry sources and a wet mixture, this task becomes reverse engineering \cite{lee2023blind, Mitcheltree_2021, guo2023automatic, colone2023reverse}.
Some fixed the graph and estimated only the parameters \cite{steinmetz2020diffmixconsole, uzrad2024diffmoog, colone2023reverse, engel2020ddsp, ramirez2021differentiable}.
Others tried to predict the graph \cite{guo2023automatic} or both \cite{lee2023blind, Mitcheltree_2021, ye2023fm, caspe2022ddx7}.
Also, with several exceptions \cite{lee2023blind, ye2023fm}, most works consider simple compositions, e.g., serial chains, that do not require graph terminology. Yet, we treat those cases as simple graphs.
Our work is the first graph reverse engineering method based on end-to-end parameter optimization with gradient descent. 

\subsection{Differentiable Signal Processing}
\looseness=-1
In the audio signal processing literature, ``differentiable" has been used as an umbrella term that refers to all methods that calculate the exact or approximate gradients of the audio processor. The most straightforward method is to implement the processors in an automatic differentiation framework, e.g., \texttt{PyTorch} \cite{paszke2019pytorch}. 
One common application of differentiable processors is to combine them with neural networks.
Thus, converting the processors to (1) be ``GPU-friendly'' and (2) provide meaningful gradients has been an active research topic \cite{colone2023reverse, steinmetz2022style, colonel2022reverse, carson2023differentiable, nercessian2020neural, lee2022differentiable, hayes2023review}.
A notable example is the sampling frequency response of an infinite impulse response (IIR) filter to sidestep the recurrent calculation and speed up the computation \cite{nercessian2020neural, lee2022differentiable}.
For processors that are not linear time-invariant (LTI), other approximation techniques have been proposed, e.g., replacing the nonlinear recurrent part with an IIR filter \cite{steinmetz2022style} or assuming frame-wise LTI to a linear time-varying system \cite{carson2023differentiable}.

If all processors are differentiable, so is the entire audio processing graph composed with them due to the chain rule.
Therefore, the only remaining practical consideration is the compute speed, which we optimize with the batched processing technique.
Finally, one might be interested in differentiation with respect to the graph structure.
The proposed method performs this to a limited extent; pruning is a binary operation that modifies the graph structure. 
We relaxed this to a continuous dry/wet weight and optimized it via gradient descent, jointly with other parameters.

\subsection{Graph Search} \label{section:related_graph_search}
\looseness=-1
Various research domains exist that search for graphs with desired properties.
For instance, neural network pruning \cite{castellano1997iterative, he2023structured} aims to find a smaller, computationally efficient sub-network that retains the performance of the full network. 
This is conceptually identical to our approach.
On the other hand, neural architecture search (NAS) aims to find a neural network architecture that achieves improved performance \cite{elsken2019neural}. 
In this case, the search space consists of graphs, with each node (or edge) representing one of the primitive neural network layers.
One particularly relevant work to ours is a differentiable architecture search (DARTS) \cite{liu2018darts}, which relaxes the choice of each layer to a categorical distribution and optimizes it via gradient descent. 
Theoretically, our method can be naturally extended to this approach; we only need to change our $2$-way choice (prune or not) to $(N+1)$-way (bypass or one of $N$ processors).
DARTS is clearly more flexible and general. 
However, it also greatly increases the computational cost, as we must compute all $N$ processors to compute their weight sum for every node. 

Another related domain is the generation/design of molecules with desired chemical properties \cite{elton2019deep}.
One dominant approach for this task is to use reinforcement learning (RL), which estimates each graph by making a sequence of decisions, e.g., adding nodes and edges \cite{you2018graph}.
RL is an attractive choice since we can be completely free of prior assumptions on graphs, and we can use arbitrary quality measures (rewards) that are not differentiable.
However, applying RL to our task has a risk of obtaining nontrivial mixing graphs that are difficult for practitioners to interpret. 
Also, it may need much larger computational resources to explore the search space sufficiently.

We note that another possible strategy for graph estimation is simply to train a neural network as a graph estimator in a supervised manner \cite{lee2023blind}. 
However, in our case, the ground-truth graphs for training are not available. Hence, they should be synthetically generated with a predefined set of rules. This risks training the model with a data distribution that is different from the real world.
Moreover, unlike the prior work \cite{lee2023blind} that only considered small graph sizes, real-world multitrack mixes typically involve significantly more source tracks and processors, making the synthesis of realistic data even more challenging. 
\section{Differentiable Processors} \label{section:processor-details}
Each processor $v_i \in V$ takes an input $u_i[n]$ and computes an output audio $y_i[n]$ (both are stereo; $n$ denotes a discrete time index) as follows: 
\begin{subequations}\label{eq:computation}
\begin{align}
    \bar{y}_i[n] &= f_i(u_i[n], \bar{p}_i), \\
    y_i[n] &=  w_i \bar{y}_i[n] + (1-w_i)u_i[n]. 
\end{align}
\end{subequations}
That is, after the main processing $f_i$ with parameters $\bar{p}_i$, we 
mix the input and the result $\bar{y}_i$ with a ``dry/wet'' weight $w_i \in [0, 1]$.
Separating the dry/wet mix from the other processing is helpful for the latter discussion on pruning. We denote the full parameter as $p_i = (\bar{p}_i, w_i)$ and use it when the distinction of the two is unnecessary.
Note that, as the processor $v_i$ is used inside the graph $G = (V, E)$, its input will be given as a sum of other nodes' outputs.
\begin{equation}
    u_i[n] = \sum_{j \in \mathcal{N}(i)} y_j[n]
\end{equation}
Here, $\mathcal{N}(i)$ denotes a set of node indices that is connected to processor $v_i$; in other words,
$e_{ji} \in E$.

\subsection{Gain/panning}
We multiply a learnable gain for each input channel.
The parameter $\bar{p}_\textttt{g} \in \mathbb{R}^2$ is the concatenation of the log gains, and the output is given as
\begin{equation}
    \bar{y}[n] = \exp (\bar{p}_\textttt{g}) \cdot u[n].     
\end{equation}

\subsection{Stereo Imager}
We adjust the stereo width by altering the loudness of the side channel (left minus right) using a single logarithmic gain parameter.
The mid and side signals are calculated as
\begin{subequations}
\begin{align}
    y_\mathrm{m}[n] &= u_\mathrm{l}[n] + u_\mathrm{r}[n], \\
    y_\mathrm{s}[n] &= \exp (\bar{p}_\textttt{s}) \cdot (u_\mathrm{l}[n] - u_\mathrm{r}[n]).
\end{align}
\end{subequations}
Then, these are converted back to the stereo output with
\begin{subequations}
\begin{align}
    \bar{y}_\mathrm{l}[n] = (y_\mathrm{m}[n]+y_\mathrm{s}[n])/2, \\
    \bar{y}_\mathrm{r}[n] = (y_\mathrm{m}[n]-y_\mathrm{s}[n])/2.
\end{align}
\end{subequations}

\subsection{Equalizer}
We use a zero-phase FIR filter with log-magnitudes used as a parameter $\bar{p}_\textttt{e}$.
We can obtain the time-domain response with exponentiation followed by inverse FFT (IFFT) and Hann-windowing $v^\mathrm{Hann}[n]$.
As a result, the length-$N$ FIR is
\begin{equation}
    h_\textttt{e}[n] = v^\mathrm{Hann}[n] \cdot \frac{1}{N} \sum_{k=0}^{N-1} \exp \bar{p}_\textttt{e}[k] \cdot  w_{N}^{kn}
\end{equation}
where $-(N+1)/2 \leq n \leq (N+1)/2$ and $w_{N} = \exp(j\cdot 2\pi/N)$. 
We convolve the same FIR to the left and right input channels to compute the output.
\begin{equation}
    \bar{y}_\mathrm{x}[n] = u_\mathrm{x}[n]*h_\textttt{e}[n] \quad (\mathrm{x} \in \{\mathrm{l}, \mathrm{r}\}).
\end{equation}
The FIR length is set to $N=2047$, i.e., the parameter $p_\textttt{e}$ is a $1024$-dimensional vector.
For all causal convolutions, \texttt{FlashFFTConv} \cite{fu2023flashfftconv} is used to speed up the processing and save memory.

\subsection{Reverb}
We employ a stereo FIR filter where each FIR is a variant of the time-varying filtered noise \cite{engel2020ddsp}. 
We start from uniform noise, $u_\mathrm{m}[n]$ and $u_\mathrm{s}[n] \sim \mathcal{U}(-1, 1)$, one for the mid channel and the other one for the side channel, respectively (both are $2$ seconds long). 
Then, each noise is converted to its STFT $U_\mathrm{x}[k, m]$ and multiplied with an independently parameterized magnitude mask $M_\mathrm{x}[k, m]$. 
\begin{equation}
    H_\mathrm{x}[k, m] = U_\mathrm{x}[k, m] \odot M_\mathrm{x}[k, m] \quad (\mathrm{x} \in \{\mathrm{m}, \mathrm{s}\}).
\end{equation}
Here, $k$ and $m$ denote frequency and time frame index, respectively.
We create each mask $M_\mathrm{x}$ with two parameters: a log-magnitude vector $H^0_\mathrm{x}[k]$ that corresponds to the initial coloration and another log-magnitude vector $H^\Delta_\mathrm{x}[k]$ that models the frequency-dependent decay of the reverb.
\begin{equation}
	M_\mathrm{x}[k, m] = \exp ({H^0_\mathrm{x}[k] + (m-1) H^\Delta_\mathrm{x}[k]}).
\end{equation}
The masked STFTs, $H_\mathrm{m}$ and $H_\mathrm{s}$, are converted back to the time-domain responses with the inverse STFT and further converted to the stereo FIR $h_\textttt{r}[n]$.
Channel-wise convolutions between the input $u[n]$ and the FIR yield the desired output $\bar{y}[n]$.
We set the FFT and hop lengths to $384$ and $192$, respectively.
Each reverb processor has $768$ parameters ($2$ channels, each with $2$ log-magnitude vectors of size $192$).

\subsection{Compressor}
We follow the canonical implementation of the digital feed-forward compressor
\cite{giannoulis2012digital}.
We first calculate a mid-channel signal $u_\mathrm{m}[n]$ by summing the left and right channels. Then, we square the signal and smooth it with the following recursive filter called ``ballistics."
\begin{equation}\label{eq:envelope}
	g_u[n] = \alpha[n] g_u[n-1]+(1-\alpha[n]) u_\mathrm{m}^2[n].
\end{equation}
In standard compressors, values of the smoothing coefficient $\alpha[n]$ are set differently for an ``attack'' and ``release'' phase (where $g_u[n]$ increases and decreases, respectively).
However, this part bottlenecks the computation speed in GPUs.
To mitigate this, we follow the recent work \cite{steinmetz2022style} and set the coefficients to a single parameter $\alpha$.
This simplification makes the process a one-pole IIR filter whose impulse response is simply given as
\begin{equation} \label{eq:ballistics-approximation}
	h^\mathrm{env}[n] = (1-\alpha) \alpha^n.
\end{equation}
Then, we can compute the energy envelope $g_u[n]$ by a simple convolution of the energy signal $\smash{u^2_\mathrm{m}[n]}$ with the above IIR $h^\mathrm{env}[n]$ truncated to length $N$.
Next, we calculate the compressed energy envelope $\smash{G_{\bar{y}}[n]}$ from the log of the input envelope $G_u[n] = \log g_u[n]$.
We use a quadratic knee that interpolates the compression and the bypass region.
For a given threshold $T$ and  half of the knee width $W$, 
\begin{equation}
	G_y[n] = \begin{cases}
		G_{\bar{y}}^\mathrm{above}[n] & G_u[n] \geq T+W,  \\
		G_{\bar{y}}^\mathrm{mid}[n]   & T-W \leq G_u[n] < T+W, \\
		G_{\bar{y}}^\mathrm{below}[n] & G_u[n] < T-W
		\end{cases}
\end{equation}
where, for a given compression ratio $R$, each term is 
\begin{subequations}
\begin{align}
	G_{\bar{y}}^\mathrm{above}[n] &= T+\frac{G_u[n]-T}{R}, \\
	G_{\bar{y}}^\mathrm{mid}[n]   &= G_u[n] + \Big(\frac{1}{R}-1\Big)\frac{(G_u[n]-T+W)^2}{4W},
\end{align}
\end{subequations}
and $G_{\bar{y}}^\mathrm{below}[n] = G_u[n]$. 
Finally, the output is given as
\begin{equation}
	{\bar{y}}_\mathrm{x}[n] = \exp(G_{\bar{y}}[n]-G_u[n]) \cdot u_\mathrm{x}[n] \quad (\mathrm{x} \in \{\mathrm{l}, \mathrm{r}\}).
\end{equation}
The scalar parameters, $\alpha$, $T$, $W$, and $R$, form the compressor's parameter vector $\bar{p}_\textttt{c} \in \mathbb{R}^4$.

\subsection{Noisegate}
Its implementation is identical to the compressor except for the output gain computation: 
$G_{\bar{y}}^\mathrm{above}[n] = G_u[n]$ and the remaining is given as
\begin{subequations}
\begin{align}
	G_{\bar{y}}^\mathrm{mid}[n]   &= G_u[n] + (1-R)\frac{(G_u[n]-T-W)^2}{4W}, \\
	G_{\bar{y}}^\mathrm{below}[n] &= T+R(G_u[n]-T).
\end{align}
\end{subequations}

\subsection{Multitap Delay}
We observed some mixes in our dataset use heavy delay (or ``echo") effects; hence, we introduce an FIR filter that can mimic this.
The left and right channels use independent FIRs, and they are convolved with the input channel-wise. For simplicity, we omit this in the following text and equation.
Under this setup, the multitap delay's FIR is given as\begin{equation}
	h_\textttt{d}[n] = \sum_{m=1}^{M} c_m[n]*\delta[n-d_m]
\end{equation}
where $\delta[n]$ is a unit impulse signal and $c_m$ is a $39$-tap FIR implemented the same as the zero-phase equalizer.
We use $M=20$ delays, and each delay $d_m$ is constrained to be at every $100\si{ms}$ range.
We want to optimize each discrete delay length $d_m \in \mathbb{N}$ with gradient descent. To achieve this, we utilize the property that each delayed impulse corresponds to a complex sinusoid in the frequency domain. A recent work \cite{hayes2023sinusoidal} shows that the complex angular frequency $z_m \in \mathbb{C}$ of such a sinusoid can be optimized using gradient descent if we allow it to be in the unit disk, i.e., $|z_m| \leq 1$.
We leverage this finding; from the angular frequency $z_m$, we compute a damped sinusoid. Then, its inverse FFT can be used as a surrogate of the delayed unit impulse.\begin{equation}
	\delta[n-d_m] \approx \frac{1}{N} \sum_{k=0}^{N-1} z_m^k w_N^{kn}.
\end{equation}
Note that this surrogate is not identical to the original delay unless the angular frequency is on the discrete bins of the unit disk. 
Therefore, we use it only for calculating the gradients with straight-through estimation \cite{bengio2013estimating}. 
Moreover, we apply normalization to the gradient and add a regularization that guides the parameters to increase their radius. 
We empirically observed that this improves the performance. 
To summarize, each gradient is updated as follows: 
\begin{equation}
	\frac{\partial L}{\partial z_m^*} \leftarrow \mathrm{sgn}\! 
	\left(
		\sum_n 
		\frac{\partial L}{\partial h_\textttt{d}[n]} 
		\frac{\partial \tilde{h}_\textttt{d}[n]}{\partial z_m^*} 
	\right)
	+ \gamma (|z_m| - 1) \;\;\!\!\mathrm{sgn} (z_m^*)
\end{equation}
where $\tilde{h}_\textttt{d}[n]$ is the surrogate FIR obtained with the continuous delay,
$\mathrm{sgn}(z) = z/|z|$ and $\gamma$ is a regularization strength set to $0.01$. 
This multitap delay has a parameter vector $\bar{p}_\textttt{d}$ of size $880$ ($20$ delays for each channel; each delay with a real and imaginary part of the sinusoid's angular frequency and $20$ log-magnitude bins of the FIR filter). 
\section{Batched Processing on Graphs}\label{section:diff_graph}
Here, we report an efficient implementation of computing the output $y$ from the source audio $\bfS$, graph $G$, and its parameters $\bfP$.
In practice, each graph $G$ is represented with a vector of node types $\bfT \in \mathbb{N}^{|V|}$ and a tensor of edge indices $\bfE \in \mathbb{N}^{2\times|E|}$. The collection of parameters $\bfP$ is implemented as a dictionary where each key-value pair corresponds to the node type $t$ and parameters $\bfP[t]\in \mathbb{R}^{|V_t|\times N_t}$ of that type with a single tensor where $|V_t|$ and $N_t$ denote the number of type-$t$ nodes and parameters, respectively.
Note that we use $\bfw$ and $\bar{\bfP}$ to denote all the collections of dry/wet weights and the remaining parameters, i.e., $\bfP = (\bar{\bfP}, \bfw)$, respectively.

\begin{figure}[t]
    \vspace{-3mm}
    \captionsetup[subfloat]{captionskip=4pt}
    \centering
    \subfloat[Target graph \label{fig:full}]{
        \includegraphics[width=.49\columnwidth]{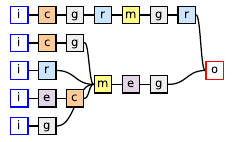}} 
    \subfloat[Optimal schedule \label{fig:oracle}]{
        \includegraphics[width=.49\columnwidth]{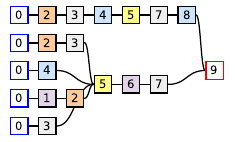}}\\
    \vspace{-1mm}
    \subfloat[Greedy schedule \label{fig:greedy}]{
        \includegraphics[width=.49\columnwidth]{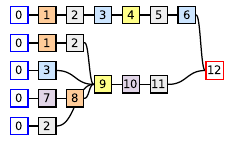}}
    \subfloat[One-by-one processing \label{fig:1by1}]{
        \includegraphics[width=.49\columnwidth]{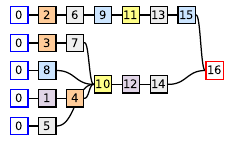}} \\
\caption{
Various schedules for batched node processing. 
For each schedule, the processing orders are shown inside the nodes. 
}
  \label{fig:schedule} 
\end{figure}

\subsection{Node Subset Sequence}
Instead of computing each node one by one \cite{uzrad2024diffmoog, engel2020ddsp}, we can batch-process multiple nodes in parallel. 
Specifically, a sequence of node subsets
$V_0, \cdots, V_N \subset V$ that satisfies the following can be used for the batched processing.
\begin{enumerate}[label=(\roman*), leftmargin=7mm]
  \setlength\itemsep{0em}
    \item 
		It is a \emph{partition}: $\cup_n V_n = V$ and $V_n\cap V_m = \emptyset$ if $n \neq m$.
    \item 
		It is \emph{causal}: there is
		no path from $u \in V_n$ to $v \in V_m$
		if $n \geq m$.
    \item 
		Each subset $V_n$ is \emph{homogeneous}: nodes with the same type $t_n$ can be in the subset.
\end{enumerate}
With this sequence, each subset's output $\bfY_n$ can be batch-computed by gathering the inputs, aggregating them if necessary, and computing the processed result.
We can further set the first and last subsets, $V_0$ and $V_N$, to collect all the input and output nodes, respectively.
Then, we can reduce the number of gather-aggregate-process iterations from $|V|$ to $N$. 
Figure \ref{fig:schedule} shows an example.
While the target graph comprises $|V|=21$ nodes (\ref{fig:full}), 
there is a node subset sequence of $N=9$ that computes $1$ equalizer \texttt{e}, $3$ compressors $\texttt{c}$, $3$ gain/pannings \texttt{g}, and so on  (\ref{fig:oracle}). 

\subsection{Type Scheduling}
\looseness=-1
To maximize the batched processing, we want to reduce the length of the subset sequence.
When the type $t_i$ of each subset $V_i$ is fixed, we always want to make the subset as maximal as possible.
With this setup, a subset sequence becomes equivalent to a ``type string," e.g., \texttt{iecgrmegro} for the previous example (\ref{fig:oracle}).
For our mixing consoles and their pruned versions, we know the optimal node sequence is a substring of \texttt{iecnsgdrmecnsgdro}.
If such information is not available, we must solve a more general scheduling problem. 
Given that the search tree for the shortest sequence grows exponentially, a brute-force search becomes computationally impractical for most graphs. 
As an alternative, a {greedy} approach can be employed, where the type with the largest number of computable nodes is selected at each step (\ref{fig:greedy}). For improved efficiency and accuracy, a more advanced beam search method can be utilized.

\begin{algorithm}[t]
\caption{Batched node processing.}\label{alg:cap}
\begin{algorithmic}[1]
\Require Types $\bfT$, edges $\bfE$, parameters $\bfP$, and inputs $\bfS$
\Ensure Output signal $y$ and regularization loss $L_\mathrm{r}$
\vspace{.7mm}
\State $\bbfT, N \gets \fn{ScheduleBatchedProcessing}(\bfT, \bfE)$ \label{algo:render:schedule}
\State $\sigma \gets \fn{OptimizeNodeOrder}(\bbfT, \bfT, \bfE)$ \label{algo:render:find-order}
\State $\bfT, \bfE, \bfP \gets \fn{Reorder}(\sigma, \bfT, \bfE, \bfP)$ \label{algo:render:reorder}
\State $\bfIG, \bfIP, \bfIA, \bfIS \gets \fn{GetReadWriteIndex}(\bbfT, \bfT, \bfE, \bfP)$ \label{algo:render:calculate_index}
\State $\bfU, L_\mathrm{r} \gets \fn{Initialize}(\bfS, \bfT), 0$ \label{algo:render:initialize}
\For{$n$ $\gets$ $1$ to $N$} \label{algo:render:iter-start}
	\State $\bbfU_n \gets \fn{Gather}(\bfU, \bfIG_n)$ 
	\Comment{\texttt{index\_select}}
	\label{algo:render:gather-input}
    \State $\bfU_n \gets \fn{Aggregate}(\bbfU_n, \bfIA_n)$ 
	\Comment{\texttt{scatter}}
	\label{algo:render:aggregate}
	\State $\bfP_n \gets \fn{Gather}(\bfP[\bar{t}_n], \bfIP_n)$ 
	\Comment{\texttt{slice}}
	\label{algo:render:gather-param}
	\State $\mathbf{Y}_n, L_{\mathrm{r}, n} \gets \fn{Process}(\bar{t}_n, \bfU_n, \bfP_n)$
	\label{algo:render:process}
	\State $\bfU \gets \fn{Store}(\mathbf{U}, \mathbf{Y}_n, \bfIS_n)$ 
	\Comment{\texttt{slice}}
	\label{algo:render:store}
        \State $L_\mathrm{r} \gets L_\mathrm{r} + L_{\mathrm{r}, n}$ \label{algo:render:reg}
\EndFor \label{algo:render:iter-end}
\State \Return $\bfY_N, L_\mathrm{r}$ \label{algo:render:out}
\end{algorithmic}
\label{algo:rendering} 
\end{algorithm}

\subsection{Implementation Details}
\looseness=-1
Algorithm \ref{algo:rendering} describes the output computation (inside the following parentheses denote the line numbers).
First, we compute a list of node types $\bbfT \in \mathbb{N}^{N+1}$, i.e., a batched node processing schedule (\ref{algo:render:schedule}).
Since the computation involves multiple memory reads/writes during 
(\ref{algo:render:iter-start}-\ref{algo:render:iter-end}), we change the node orderings to maximize contiguous memory access.
(\ref{algo:render:find-order})
This procedure allows memory access via \texttt{slice}; 
see the comments of Algorithm \ref{algo:rendering}.
After updating the graph tensors with the new order $\sigma$ (\ref{algo:render:reorder}), we retrieve lists of indices for the tensor read/writes, $\bfIG$, $\bfIP$, $\bfIA$, and $\bfIS$ (\ref{algo:render:calculate_index}).
These preprocessing steps (\ref{algo:render:schedule}-\ref{algo:render:calculate_index})
can be done on a CPU with multiple separate threads, i.e., they do not consume the GPU compute and memory.
After the preprocessing, we create a tensor 
$\bfU$ with a shape $|V|\times 2\times L$ that stores all the node outputs from the processors. 
Since all the inputs are in the first partition $V_0$, this tensor is initialized as $\bfU = \bfS \oplus \mathbf{0}$ where $\oplus$ denotes concatenation.
A regularization loss $L_\mathrm{r}$ that aggregates each iteration's losses is also initialized (\ref{algo:render:initialize}).

The remaining is the main loop, repeating the processor output computation and necessary reads/writes 
(\ref{algo:render:iter-start}-\ref{algo:render:iter-end}).
During each $n^\mathrm{th}$ iteration, we gather the previous outputs $\bbfU_n$ routed to the current partition nodes by accessing the intermediate tensor $\bfU$ using the index $\bfIG_n$ with the \texttt{index\_select} operation (\ref{algo:render:gather-input}). If needed, these outputs are aggregated using \texttt{scatter} (\ref{algo:render:aggregate}). Similarly, the parameter tensor $\bfP_n$ is retrieved using the corresponding index $\bfIP_n$; our node reordering strategy (\ref{algo:render:reorder}) simplifies this operation to a \texttt{slice}, which is more efficient than the usual \texttt{index\_select}.
With the input audio $\bfU_n \in \mathbb{R}^{|V_n|\times 2\times L}$ and parameters $\bfP_n \in \mathbb{R}^{|V_n|\times N_t}$ obtained, we batch-process the node outputs $\bfY_n$ (\ref{algo:render:process}). These outputs are then stored back into the intermediate tensor $\bfU$ using the \texttt{slice} index $\bfIS_n$ (\ref{algo:render:store}), ensuring they are accessible for subsequent steps. Additionally, we add the intermediate regularization loss $L_{\mathrm{r}, n}$ in the computation (\ref{algo:render:reg}).
At the end of the iteration, all node outputs are saved in $\bfU$. The final outputs of the graph are given by the outputs of the last step, $y = \bfY_N \in \mathbb{R}^{1\times 2\times L}$, as the final node partition $V_N$ has the output node (\ref{algo:render:out}).
\section{Music Mixing Consoles}\label{section:mixing_consoles}
Before exploring the pruning, we first evaluate the match quality of the mixing consoles, which will serve as a performance upper bound.
In addition to evaluating the full console, we analyze the contribution of each processor type to the overall performance. To achieve this, we begin with a base graph that simply sums all the inputs. Then, we sequentially add each processor type to every processor chain (refer to Tables \ref{table:mixing-console-medley} and \ref{table:mixing-console-mixing-secrets}). Each preliminary graph is then optimized and evaluated for every song in the validation set.
This experiment is similar to one conducted in the recent work \cite{colone2023reverse}, albeit the structure of mixing consoles and the processors differ.

\subsection{Optimization}
The graph parameters and dry/wet weights are optimized with the following audio-domain loss,
\begin{equation}
    L_\mathrm{a} = \alpha_{\mathrm{lr}}L_\mathrm{lr} + \alpha_{\mathrm{m}}L_\mathrm{m} + \alpha_{\mathrm{s}}L_\mathrm{s},
\end{equation}
where each term $L_\mathrm{x}$ is a multi-resolution STFT (MRSTFT) loss \cite{yamamoto2020parallel} ($\mathrm{x \in \{lr, m, s\}}$, $\mathrm{lr}$: left/right, $\mathrm{m}$: mid, $\mathrm{s}$: side)
\begin{equation}
    L_\mathrm{x} = \sum_{i=1}^I \Bigg[ 
		\frac{\| \log Y^{(i)}_\mathrm{x} - \log \hat{Y}^{(i)}_\mathrm{x} \|_1}{N} 
	    + \frac{\|Y^{(i)}_\mathrm{x} - \hat{Y}^{(i)}_\mathrm{x} \|_F}{\|Y^{(i)}_\mathrm{x}\|_F} 
		\Bigg].
\end{equation}
Here, $Y_\mathrm{x}^{(i)}$ and $\hat{Y}_\mathrm{x}^{(i)}$ represent the $i^\mathrm{th}$ Mel spectrograms of the target and predicted mix, respectively. The terms $N$, $\|\cdot\|_1$, and $\|\cdot\|_F$ refer to the number of frames, the $l_1$ norm, and the Frobenius norm, respectively. FFT sizes of $512$, $1024$, and $4096$ are used, with hop sizes set to $1/4$ of their respective FFT sizes. The number of Mel filterbanks is $96$ for all scales.
Before each STFT, A-weighting is applied \cite{wright2020perceptual}. The loss weights are $\alpha_\mathrm{lr} = 0.5$, $\alpha_\mathrm{m}=0.25$, and $\alpha_\mathrm{s}=0.25$. We use \texttt{auraloss} \cite{steinmetz2020auraloss} for the implementation. Additionally, we include the following regularization term to encourage gain-staging, a common practice to ensure the total energy of the input and output remains approximately equal:
\begin{equation}
    L_\mathrm{g} = \sum\nolimits_{v_i \in V_\mathrm{g}} \left| \log \| (\bar{y}_i)_\mathrm{m} \|_2 - \log \| (u_{i})_{\mathrm{m}} \|_2 \right|
\end{equation}
where $(\cdot)_\mathrm{m}$ and $\|\cdot\|_2$ denote mid channel and $l_2$ norm, respectively.
This regularization is applied to all equalizers, reverbs, and multitap delays (denoted as $V_\mathrm{g}$).
This approach enables us to remove redundant gains introduced by these LTI processors and constrain the parameters to be within a reasonable range.
Therefore, the full optimization loss is 
\begin{equation}
    L(\bar{\mathbf{P}}, \mathbf{w}) = L_\mathrm{a}(\bar{\mathbf{P}}, \mathbf{w}) + \alpha_\mathrm{g} L_\mathrm{g}(\mathbf{\bar{P}})
\end{equation}
where the weight of the gain-staging regularization is set to $\alpha_\mathrm{g} = 10^{-3}$ (a different notation was used from Equation \ref{eq:objective} to emphasize the optimized parameters).
Parameters are initialized with a Gaussian distribution $\mathcal{N}(0, 10^{-2})$. For dry/wet weights, a logistic sigmoid is further applied to center the initial values to $0.5$ and restrict their range to $(0, 1)$.
We optimize each console for $12\si{k}$ steps using AdamW optimizer \cite{loshchilov2017decoupled} 
with a $0.01$ learning rate.
Each step randomly samples a $3.8\si{s}$ segment, computes the mix $\hat{y}$, 
and compares it with the corresponding ground truth $y$ (hence, the batch size is $1$). The first second of each mix is omitted when calculating the loss; it is used only to ``warm up'' the processors with long states, e.g., compressors and reverbs. 

\begin{table*}[!ht]
\setlength\tabcolsep{2.7pt}
\renewcommand{\arraystretch}{.85}
\begin{center}
\caption{
Matching performances of the mixing consoles using different processor type configurations. Dataset: \texttt{MedleyDB}.
}
\small
\begin{tabular}{lcccccccccccc}
\toprule
& 
& \multicolumn{4}{c}{MRSTFT $\downarrow$} 
& \multirow{2}[1]{*}{FAD $\downarrow$}
& \multirow{2}[1]{*}{SI-SDR $\uparrow$} 
& 
\multicolumn{5}{c}{MIR Feature Distance $\downarrow$}\\
\cmidrule(r){3-6}
\cmidrule(l){9-13}
& 
 & $L_\mathrm{a}$ & $L_\mathrm{lr}$ & $L_\mathrm{m}$ & $L_\mathrm{s}$  & & & $d_\mathrm{RMS}$ & $d_\mathrm{CR}$ & $d_\mathrm{SW}$ & $d_\mathrm{SI}$ & $d_\mathrm{BS}$ 
\\ 
\midrule
\arrayrulecolor{lightgray}
\multicolumn{2}{l}{Base graph (sum of dry sources)}
& $50.7$ & $1.42$ & $1.42$ & $198$ & $3.57$ & $-6.41$ & $-4.68$ & $.662$ & $-1.99$ & $-2.22$ & $2.46$ \\
\midrule
+ Gain/panning 
& \texttt{\textcolor{white}{ecns\textcolor{black}{g}dr}}
& $.550$ & $.583$ & $.485$ & $.549$ & $.719$ & $-7.20$ & $-5.01$ & $.484$ & $-3.52$ & $-2.38$ & $-1.56$ \\
+ Stereo imager 
& \texttt{\textcolor{white}{ecn\textcolor{black}{sg}dr}}
& $.541$ & $.564$ & $.483$ & $.553$ & $.733$ & $-6.97$ & $-5.02$ & $.488$ & $-3.54$ & $-2.48$ & $-1.54$ \\
+ Equalizer 
& \texttt{\textcolor{white}{\textcolor{black}{e}cn\textcolor{black}{sg}dr}}
& $.450$ & $.453$ & $.390$ & $.504$ & $.573$ & $-5.25$ & $-5.45$ & $.415$ & $-3.71$ & $-2.65$ & $-1.62$ \\
+ Reverb 
& \texttt{\textcolor{white}{\textcolor{black}{e}cn\textcolor{black}{sg}d\textcolor{black}{r}}}
& $.368$ & $.361$ & $.360$ & $.390$ & $.405$ & $-3.94$ & $-5.54$ & $.356$ & $-4.17$ & $-3.55$ & $-1.66$ \\
+ Compressor 
& \texttt{\textcolor{white}{\textcolor{black}{ec}n\textcolor{black}{sg}d\textcolor{black}{r}}}
& $.315$ & $.304$ & $.297$ & $.356$ & $.312$ & $-2.95$ & $-6.15$ & $.173$ & $-4.33$ & $-3.61$ & $-1.52$\\
+ Noisegate 
& \texttt{\textcolor{white}{\textcolor{black}{ecnsg}d\textcolor{black}{r}}}
& $.302$ & $.288$ & $.281$ & $.353$ & $.264$ & $-2.60$ & $-6.12$ & $.065$ & $-4.22$ & $-3.55$ & $-1.51$ \\
\midrule
+ Multitap delay (full) 
& \texttt{\textcolor{white}{\textcolor{black}{ecnsgdr}}}
& $.296$ & $.288$ & $.284$ & $.324$ & $.222$ & $-2.66$ & $-6.17$ & $.093$ & $-4.34$ & $-3.78$ & $-1.46$ \\
\arrayrulecolor{black}
\bottomrule
\end{tabular}

\label{table:mixing-console-medley}
\end{center}
\end{table*}

\begin{table*}[!ht]
\setlength\tabcolsep{2.7pt}
\renewcommand{\arraystretch}{.85}
\begin{center}
\caption{
Matching performances of the mixing consoles using different processor type configurations. Dataset: \texttt{MixingSecrets}.
}
\small
\begin{tabular}{lcccccccccccc}
\toprule
& 
& \multicolumn{4}{c}{MRSTFT $\downarrow$} 
& \multirow{2}[1]{*}{FAD $\downarrow$}
& \multirow{2}[1]{*}{SI-SDR $\uparrow$} 
& 
\multicolumn{5}{c}{MIR Feature Distance $\downarrow$}\\
\cmidrule(r){3-6}
\cmidrule(l){9-13}
& 
 & $L_\mathrm{a}$ & $L_\mathrm{lr}$ & $L_\mathrm{m}$ & $L_\mathrm{s}$  & & & $d_\mathrm{RMS}$ & $d_\mathrm{CR}$ & $d_\mathrm{SW}$ & $d_\mathrm{SI}$ & $d_\mathrm{BS}$ 
\\ 
\midrule
\arrayrulecolor{lightgray}
\multicolumn{2}{l}{Base graph (sum of dry sources)}
& $7.30$ & $2.16$ & $2.02$ & $22.9$ & $2.78$ & $-13.8$ & $-4.16$ & $1.15$ & $-1.66$ & $-2.22$ & $.401$ \\
\midrule
+ Gain/panning 
& \texttt{\textcolor{white}{ecns\textcolor{black}{g}dr}}
& $.876$ & $.856$ & $.819$ & $.973$ & $1.49$ & $-10.0$ & $-4.63$ & $1.21$ & $-2.33$ & $-1.82$ & $-.842$ \\
+ Stereo imager 
& \texttt{\textcolor{white}{ecn\textcolor{black}{sg}dr}}
& $.847$ & $.834$ & $.791$ & $.928$ & $1.46$ & $-10.3$ & $-4.64$ & $1.20$ & $-2.27$ & $-1.94$ & $-.855$ \\
+ Equalizer 
& \texttt{\textcolor{white}{\textcolor{black}{e}cn\textcolor{black}{sg}dr}}
& $.699$ & $.698$ & $.622$ & $.780$ & $.933$ & $-9.62$ & $-5.00$ & $1.14$ & $-2.79$ & $-2.20$ & $-1.23$ \\
+ Reverb 
& \texttt{\textcolor{white}{\textcolor{black}{e}cn\textcolor{black}{sg}d\textcolor{black}{r}}}
& $.614$ & $.601$ & $.579$ & $.674$ & $.642$ & $-6.89$ & $-5.07$ & $1.04$ & $-2.98$ & $-2.74$ & $-1.39$ \\
+ Compressor 
& \texttt{\textcolor{white}{\textcolor{black}{ec}n\textcolor{black}{sg}d\textcolor{black}{r}}}
& $.558$ & $.542$ & $.512$ & $.637$ & $.526$ & $-6.10$ & $-5.13$ & $.987$ & $-2.81$ & $-2.70$ & $-1.20$\\
+ Noisegate 
& \texttt{\textcolor{white}{\textcolor{black}{ecnsg}d\textcolor{black}{r}}}
& $.548$ & $.532$ & $.502$ & $.625$ & $.523$ & $-7.04$ & $-4.98$ & $.927$ & $-2.61$ & $-2.60$ & $-1.17$ \\
\midrule
+ Multitap delay (full) 
& \texttt{\textcolor{white}{\textcolor{black}{ecnsgdr}}}
& $.545$ & $.529$ & $.502$ & $.618$ & $.506$ & $-6.85$ & $-4.98$ & $.985$ & $-2.84$ & $-2.66$ & $-1.13$ \\
\arrayrulecolor{black}
\bottomrule
\end{tabular}

\label{table:mixing-console-mixing-secrets}
\end{center}
\end{table*}

\subsection{Evaluation Metrics}\label{section:eval-metrics}

We follow the recent work \cite{vanka2024diff} and compare the optimized mix and the ground truth with the following metrics.
First, we measure Fréchet Audio Distance (FAD) \cite{kilgour2018fr}, which calculates the distance between the feature distributions obtained with a pre-trained audio encoder. We also report the scale-invariant signal-to-distortion ratio (SI-SDR), widely used in source separation literature, which involves sample-wise comparison \cite{le2019sdr}.
Furthermore, we measure the mean square errors (MSEs) of simple analysis features used in music information retrieval (MIR). These are referred to as ``MIR feature distances."
We split the mix and its match into
disjoint $8$-second segments, $y^{(1)}, \cdots, y^{(M)}$ and $\hat{y}^{(1)}, \cdots, \hat{y}^{(M)}$. Then, we calculate the mean square errors (MSEs) of features extracted from those segments. 
We further apply a logarithm to compress the values, as we empirically found it to correlate better with the subjective similarity.
We consider the same set of features in \cite{vanka2024diff}: root mean square (RMS), crest factor (CF), stereo width (SW), stereo imbalance (SI), and bark spectrum (BS) \cite{man2014analysis}.
\begin{equation}\label{eq:mir-metric}
    d_\mathrm{x} = \log_{10} \left[\frac{1}{MK_\mathrm{x}}\sum^{M}_{m=1}\left\| F_\mathrm{x}(y^{(m)}) - F_\mathrm{x}(\hat{y}^{(m)})  \right\|_2^2 \right].
\end{equation}
where $\mathrm{x} \in \{\mathrm{RMS, CF, SW, SI, BS}\}$, $\|\cdot\|_2$ denotes $l_2$ norm, and $K_\mathrm{x}$ denotes the size of the feature $\mathrm{x}$. Each feature reflects certain aspects of the mix. For example, the CF is defined as a ratio between the peak absolute value and the RMS of the waveform. Therefore, it could be used to compare the mixes and matches in terms of loudness dynamics. 

\subsection{Data}
For each song, we need dry sources $\bfS$ and a mixture $y$ to run the optimization. Also, we need additional subgroup information to construct the initial mixing console.
Thus, we utilize the \texttt{MedleyDB} dataset \cite{bittner2014medleydb, bittner2016medleydb}, as it provides all the necessary components.
The \texttt{MixingSecrets} library \cite{senior2018mixing} only provides the audio; therefore, we manually create the subgroup information. 
For evaluation purposes, we use a random $24$-song subset for each dataset, and the results are reported separately.
Finally, to collect the graph data, we use the full data and also include our private \texttt{Internal} dataset crowdsourced from multiple engineers, which comprises Western music recordings. 
A total of $1129$ songs were collected ($188$, $472$, and $579$ songs for \texttt{MedleyDB}, \texttt{MixingSecrets}, and \texttt{Internal}, respectively).
Every dry source track and mix is stereo and resampled to $30\si{kHz}$ sampling rate.

\subsection{Results}\label{section:console-results}
Table \ref{table:mixing-console-medley} and \ref{table:mixing-console-mixing-secrets} report the results for the \texttt{MedleyDB} and \texttt{MixingSecrets} evaluation subsets, respectively.
We first report the \texttt{MedleyDB} results. 
First, the base graph reports an MRSTFT loss $L_\mathrm{a}$ of $50.7$. 
The side-channel term $L_\mathrm{s}$ is especially large, reporting $198$, as most source tracks are mono (hence, so is the naive sum) while the target mixes have wide stereo images. 
With the gain/pannings and stereo imagers, 
we achieve ``rough mixes'' with a loss of $0.541$.
Then, we fill in the missing details by adding the remaining processor types. Each type reduces the audio loss, with the full mixing console achieving a value of $0.296$.
The same trend can be observed in the \texttt{MixingSecrets} results, albeit the audio loss is higher ($0.545$ for the full console), indicating that mixes from this dataset are more challenging than \texttt{MedleyDB}'s.

In terms of other metrics, FAD is the only one that shows the same consistent improvement for each processor type.
The SI-SDR is slightly inconsistent, and the values are very low even with the full console ($-4.50 \si{dB}$). We suspect two causes: the MRSTFT loss does not model phases, and the dry tracks and target mix are not precisely aligned for some songs.
The MIR feature distances are also inconsistent, but each processor tends to improve its relevant metrics. 
While the compressor and noisegate reduce the CF distance $d_\mathrm{CR}$ ($-0.183$ and $-0.108$, respectively), they rather increase the BS distance $d_\mathrm{BS}$ ($+0.14$ and $+0.01$, respectively, on \texttt{MedleyDB}), 
possibly due to the properties of these processors that can modify the loudness dynamics, not the overall spectral balance.
Similarly, as multitap delay has two FIRs independently parameterized for the left and right channels, they are effective for matching the stereo image, which is shown by the decreased SW distance $d_\mathrm{SW}$ and SI distance $d_\mathrm{SI}$ ($-0.12$ and $-0.23$, respectively). However, it has a risk of ``smearing'' the transient and harming the loudness dynamics (an increase of $+0.028$ in the CF distance $d_\mathrm{CR}$). 
Nevertheless, we note that the average of all MIR feature distances on the two subsets consistently decreases, from $-1.22$ to $-2.63$.%

We can conduct further qualitative analysis by comparing log-magnitude spectrograms of the targets, mixing console matches, and their errors (see the top $8$ rows of Figure \ref{fig:spec-main}). 
Again, we observe that adding each processor type improves the match. 
Furthermore, each song has different types that are more crucial; for the song \texttt{NewSkin} (Figure \ref{fig:spec-main-full-a}), there is significant error reduction when reverbs are introduced. However, this is not the case for  \texttt{RockSteady} (Figure \ref{fig:spec-main-full-b}). Instead, for the latter, different from the average trend, the multitap delays improve the match more than the reverbs. 
This is not so surprising, as the original mix introduces heavy delay effects.
\section{Music Mixing Graph Search}\label{section:search}
The previous full mixing console $G_\mathrm{c}$ serves as an upper bound for matching performance. Now, our goal is to identify a sparser graph that maintains a similar match quality. This is achieved by pruning the console as much as possible while ensuring that the loss increase remains within a pre-defined tolerance threshold $\tau$.
This requirement is illustrated as a circle in Figure \ref{fig:framework-full}. We write this objective as\begin{equation}\label{eq:pruning}
    \mathrm{minimize} \;\; | V_\mathrm{p} |  \quad \mathrm{s.t.} \;\; 
	\min L_\mathrm{a} (G_\mathrm{p}) \leq \min L_\mathrm{a}(G_\mathrm{c}) + \tau
\end{equation}
where $V_\mathrm{p}$ represents the pruned graph's node set and $|\cdot|$ denotes its cardinality. The term $\min (\cdot)$ emphasizes that our focus is on the optimized audio loss after pruning. 
Note that only processors are subject to pruning, while auxiliary nodes remain unchanged.
Instead of the number of remaining nodes $|\Vp|$, we will report a pruning ratio defined as the number of pruned processors over the number of the initial console's processors. 

\subsection{Iterative Pruning}
\begin{algorithm}[t]
\caption{
Mixing graph search with iterative pruning.
}
\begin{algorithmic}[1]
\Require A mixing console $\Gc$, dry tracks $\bfS$, and mixture $y$
\Ensure Pruned graph $\Gp$ and parameters $\bfP = (\bar{\bfP}, \bfw)$
\vspace{.7mm}

\State $\bar{\bfP}, \bfw \gets \fn{Initialize}(\Gc)$ 
\label{algo:prune:init}

\State $\bar{\bfP}, \bfw \gets \fn{Train}(\Gc, \bar{\bfP}, \bfw, \bfS, y)$

\State $\La^\mathrm{min} \gets \fn{Evaluate}(\Gc, \bar{\bfP}, \bfw, \bfS, y)$ 
\label{algo:prune:console-eval}

\State $\Gp \gets \Gc$

\For{$n$ $\leftarrow$ $1$ to $N_\mathrm{iter}$} \label{algo:prune:iter-start}
	\color{blue}
	\State $V_\mathrm{pool}, \bfm \gets \fn{GetAllProcessors}(V), \mathbf{1}$ \label{algo:prune:preprocess}
 	\color{purple}
	\While {$\mathrm{TryPrune}\:\!(V_\mathrm{pool}, \bfw, \bfm)$} \label{algo:prune:while} 
  	\color{red}
		\State $V_\mathrm{cand}, {\bfm}_{\:\!\mathrm{cand}} \gets \fn{SampleCandidate}(V_\mathrm{pool}, \bfw)$ \label{algo:prune:sample}
    	\color{black}
		\State $\La \gets \fn{Evaluate}(\Gp, \bar{\bfP}, \bfw\odot \bfm \odot {\bfm}_{\:\!\mathrm{cand}}, \bfS, y)$ \label{algo:prune:try}
		\If {$\La < \La^\mathrm{min} + \tau$} \label{algo:prune:trial-if}
			\State $\La^\mathrm{min} \gets \min (\La^\mathrm{min}, \La) $ \label{algo:prune:update}
                \State $\bfm \gets \bfm \odot {\bfm}_{\:\!\mathrm{cand}}$
		\EndIf \label{algo:prune:trial-endif}
		\color{magenta}
		\State $V_\mathrm{pool} = \fn{UpdatePool}(V_\mathrm{pool}, V_\mathrm{cand})$ \label{algo:prune:update-pool}
  \color{black}
    \EndWhile
	\State $\Gp, \bar{\bfP}, \bfw \gets \fn{Prune}(\Gp, \bar{\bfP}, \bfw, \bfm)$ \label{algo:prune:actual-prune}
    \State $\bar{\bfP}, \bfw \gets \fn{Train}(\Gp, \bar{\bfP}, \bfw, \bfS, y)$
\EndFor \label{algo:prune:iter-end}
\State \Return $\Gp, \bar{\bfP}, \bfw$
\end{algorithmic}
\label{algo:pruning} 
\end{algorithm}

\begin{algorithm}[t]
\caption{
Mixing graph search using dry/wet method.
}
\begin{algorithmic}[1]
\Require A mixing console $\Gc$, dry tracks $\bfS$, and mixture $y$
\Ensure Pruned graph $\Gp$ and parameters $\bfP = ({\bar{\bfP}}, \bfw)$
\vspace{.7mm}

\State $\bar{\bfP}, \bfw \gets \fn{Initialize}(\Gc)$ 
\State $\bar{\bfP}, \bfw \gets \fn{Train}(\Gc, \bar{\bfP}, \bfw, \bfS, y)$
\State $\La^\mathrm{min} \gets \fn{Evaluate}(\Gc, \bar{\bfP}, \bfw, \bfS, y)$ 

\State $\Gp \gets \Gc$

\For{$n$ $\leftarrow$ $1$ \textbf{to} $N_\mathrm{iter}$} 
	\color{blue}
	\State $\Tc, \bfm \gets \fn{GetProcessorTypeSet}(V), \mathbf{1}$ \label{algo:prune-weight:init-start}
	\For{$t$ \textbf{in} $\Tc$}
		\State $V_t, \bfw_t \gets \fn{Filter}(V, t), \fn{Filter}(\bfw, t)$
		\State $N_t, r_t \gets |V_t|, 0.1$
	\EndFor \label{algo:prune-weight:init-end}
	\color{black}
	\color{purple}
	\While {$\Tc\neq\emptyset$} \label{algo:prune-weight:while}
	\color{red}
		\State $t \gets \fn{SampleType}(\Tc)$ \label{algo:prune-weight:sample-start}
		\State $\bar{V}_t, \bar{\bfm} \gets 
		\fn{GetLeastWeightNodes}(V_t, \bfw_t, \lfloor N_t r_t \rceil)$\!\label{algo:prune-weight:sample-end}
		\color{black}
		\State $\La \gets \fn{Evaluate}(\Gp, \bar{\bfP}, \bfw\odot \bfm \odot \bar{\bfm}, \bfS, y)$ 
		\If {$\La < \La^\mathrm{min} + \tau$} 
			\State $\La^\mathrm{min} \gets \min (\La^\mathrm{min}, \La) $ 
            \State $\bfm \gets \bfm \odot \bar{\bfm}$
			\color{magenta}
			\State $V_t \gets V_t \setminus \bar{V}_t$ \label{algo:prune-weight:update-success}
			\color{black}
		\Else
			\color{magenta}
			\If {$\lfloor N_t r_t \rceil > 1$} \label{algo:prune-weight:update-fail-start}
				\State $r_t \gets r_t/2$
			\Else
				\State $\Tc \gets \Tc \setminus \{t\}$
			\EndIf
			\color{black}
		\EndIf \label{algo:prune-weight:update-fail-end}
    \EndWhile
	\color{black}
	\State $\Gp, \bar{\bfP}, \bfw \gets \fn{Prune}(\Gp, \bar{\bfP}, \bfw, \bfm)$ 
    \State $\bar{\bfP}, \bfw \gets \fn{Train}(\Gp, \bar{\bfP}, \bfw, \bfS, y)$
\EndFor 
\State \Return $\Gp, \bar{\bfP}, \bfw$
\end{algorithmic}
\label{algo:pruning-weight} 
\end{algorithm}

Finding the optimal solution $\Vp^*$, a set with the smallest number of nodes, is computationally prohibitive due to the combinatorial nature of the problem, i.e., the number of candidate graphs is a power of the initial mixing console. 

To mitigate this, we assume that the processors independently contribute to the match and adopt a greedy pruning strategy. Following the iterative approach \cite{castellano1997iterative}, we incrementally remove processors whenever the tolerance condition is met. Ideally, intermediate pruned graphs should be fine-tuned before assessing the tolerance condition.
However, for a reasonable computational complexity, we omit this step, paying the cost of potentially missing some removable processors. 

Our method is described in Algorithm \ref{algo:pruning}, with specific steps referenced in the following parentheses.
We start from the previously described optimization of the mixing console $\Gc = (V, E)$ and evaluate the final loss (\ref{algo:prune:init}-\ref{algo:prune:console-eval}). 
The loss $\La^\mathrm{min}$ acts as a threshold, with the tolerance $\tau$ defining acceptable degradation. 
Next, we alternate between pruning and optimizing the remaining parameters and weights, i.e., fine-tuning (\ref{algo:prune:iter-start}-\ref{algo:prune:iter-end}).
Each pruning stage involves multiple trials, where subsets of candidate processors ${V}_\mathrm{cand}$ are sampled from the remaining pool $V_\mathrm{pool}$ (\ref{algo:prune:sample}) and tested for removability (\ref{algo:prune:trial-if}). If the result satisfies the constraint, pruning is applied; otherwise, it is reverted (\ref{algo:prune:trial-if}-\ref{algo:prune:trial-endif}). This process is repeated until the terminal condition (\ref{algo:prune:while}) is met.

In implementation, pruning trials are simulated by multiplying the weight vector $\bfw$ with binary masks $\bfm$ and ${\bfm}_{\:\!\mathrm{cand}}$  (\ref{algo:prune:try}). 
After the trials, we perform the actual pruning of nodes so that we can discard the unnecessary processing and accelerate further optimization (\ref{algo:prune:actual-prune}).
Note that pruning occasionally improves the match quality. In such cases, the pruning threshold is updated accordingly (\ref{algo:prune:update}).

\begin{table*}[ht]
\setlength\tabcolsep{2.7pt}
\renewcommand{\arraystretch}{.85}
\begin{center}
\caption{
Pruning results with different strategies and tolerance $\tau$. Dataset: \texttt{MedleyDB}.
}
\small
\begin{tabular}{llcccccccccccccccc}
\toprule
& & \multirow{1}[5]{*}{$L_\mathrm{a}$ $\downarrow$}
& \multicolumn{8}{c}{Pruning Ratio $\uparrow$}
& \multirow{1}[5]{*}{FAD $\downarrow$}
& \multirow{1}[5]{*}{SI-SDR $\uparrow$}
& \multicolumn{5}{c}{MIR Feature Distance $\downarrow$}
\\
\cmidrule{4-11}
\cmidrule{14-18}

& \multicolumn{1}{c}{$\tau$} & 
& $\mu$ 
& \texttt{g} 
& \texttt{s} 
& \texttt{e} 
& \texttt{r} 
& \texttt{c} 
& \texttt{n} 
& \texttt{d} 
& & & $d_\mathrm{RMS}$ & $d_\mathrm{CR}$ & $d_\mathrm{SW}$ & $d_\mathrm{SI}$ & $d_\mathrm{BS}$ 
\\

\midrule
\arrayrulecolor{lightgray}
Mixing console
& \multicolumn{1}{c}{$-$}
& $.296$ 
& $-$ 
& $-$ & $-$ & $-$ & $-$ & $-$ & $-$ & $-$ 
& $.222$ & $-2.66$ & $-6.17$ & $.093$ & $-4.34$ & $-3.78$ & $-1.46$\\

\midrule
Brute-force & $.01$ 
& $.305$ & $.626$ & $.343$ & $.811$ & $.540$ & $.772$ & $.669$ & $.711$ & $.532$
& $.262$ & $-3.41$ & $-5.94$ & $.147$ & $-4.14$ & $-3.54$ & $-1.51$
\\

Dry/wet & $.01$ 
& $.302$ & $.561$ & $.315$ & $.791$ & $.416$ & $.736$ & $.568$ & $.669$ & $.433$
& $.229$ & $-1.87$ & $-6.02$ & $.066$ & $-4.23$ & $-3.62$ & $-1.55$
\\

\midrule
\multirow{3}{*}{Hybrid} & $.001$ \;
& $.295$ & $.442$ & $.216$ & $.728$ & $.259$ & $.612$ & $.503$ & $.540$ & $.237$
& $.227$ & $-2.18$ & $-6.11$ & $.108$ & $-4.36$ & $-3.77$ & $-1.51$
\\

& $.01$   
& $.301$ & $.606$ & $.322$ & $.805$ & $.483$ & $.741$ & $.666$ & $.708$ & $.519$
& $.247$ & $-3.03$ & $-5.97$ & $.096$ & $-4.15$ & $-3.58$ & $-1.52$ \\

& $.1$  
& $.374$ & $.829$ & $.586$ & $.926$ & $.828$ & $.918$ & $.804$ & $.840$ & $.898$
& $.406$ & $-4.38$ & $-5.47$ & $.335$ & $-3.78$ & $-3.07$ & $-1.57$ \\

\arrayrulecolor{black}
\bottomrule
\end{tabular}
\label{table:pruning-medley}
\end{center}
\end{table*}

\begin{table*}[!ht]
\setlength\tabcolsep{2.7pt}
\renewcommand{\arraystretch}{.85}
\begin{center}
\caption{
Pruning results with different strategies and tolerance $\tau$. Dataset: \texttt{MixingSecrets}.
}
\small
\begin{tabular}{llcccccccccccccccc}
\toprule
& & \multirow{1}[5]{*}{$L_\mathrm{a}$ $\downarrow$}
& \multicolumn{8}{c}{Pruning Ratio $\uparrow$}
& \multirow{1}[5]{*}{FAD $\downarrow$}
& \multirow{1}[5]{*}{SI-SDR $\uparrow$}
& \multicolumn{5}{c}{MIR Feature Distance $\downarrow$}
\\
\cmidrule{4-11}
\cmidrule{14-18}

& \multicolumn{1}{c}{$\tau$} & 
& $\mu$ 
& \texttt{g} 
& \texttt{s} 
& \texttt{e} 
& \texttt{r} 
& \texttt{c} 
& \texttt{n} 
& \texttt{d} 
& & & $d_\mathrm{RMS}$ & $d_\mathrm{CR}$ & $d_\mathrm{SW}$ & $d_\mathrm{SI}$ & $d_\mathrm{BS}$ 
\\

\midrule
\arrayrulecolor{lightgray}
Mixing console
& \multicolumn{1}{c}{$-$}
& $.545$ 
& $-$ 
& $-$ & $-$ & $-$ & $-$ & $-$ & $-$ & $-$ 
& $.506$ & $-6.85$ & $-4.98$ & $.985$ & $-2.84$ & $-2.66$ & $-1.13$\\

\midrule
Brute-force & $.01$ 
& $.566$ & $.654$ & $.498$ & $.823$ & $.429$ & $.710$ & $.605$ & $.766$ & $.748$
& $.548$ & $-7.13$ & $-5.06$ & $.991$ & $-2.83$ & $-2.54$ & $-1.19$
\\

Dry/wet & $.01$ 
& $.561$ & $.566$ & $.472$ & $.818$ & $.222$ & $.623$ & $.539$ & $.744$ & $.546$
& $.513$ & $-6.79$ & $-4.96$ & $1.02$ & $-2.68$ & $-2.48$ & $-1.13$
\\

\midrule
\multirow{3}{*}{Hybrid} & $.001$ \;
& $.550$ & $.433$ & $.345$ & $.765$ & $.134$ & $.421$ & $.429$ & $.552$ & $.383$
& $.523$ & $-6.66$ & $-4.94$ & $1.00$ & $-2.68$ & $-2.58$ & $-1.11$
\\

& $.01$   
& $.563$ & $.622$ & $.490$ & $.848$ & $.344$ & $.637$ & $.599$ & $.789$ & $.646$
& $.519$ & $-6.35$ & $-5.05$ & $.995$ & $-2.78$ & $-2.61$ & $-1.22$\\

& $.1$  
& $.647$ & $.843$ & $.703$ & $.914$ & $.751$ & $.864$ & $.831$ & $.927$ & $.909$
& $.805$ & $-7.98$ & $-4.82$ & $1.03$ & $-2.78$ & $-2.33$ & $-1.19$ \\

\arrayrulecolor{black}
\bottomrule
\end{tabular}
\label{table:pruning-mixing-secrets}
\end{center}
\end{table*}

\begin{figure*}
    \begin{center}{\includegraphics[width=2.075\columnwidth]{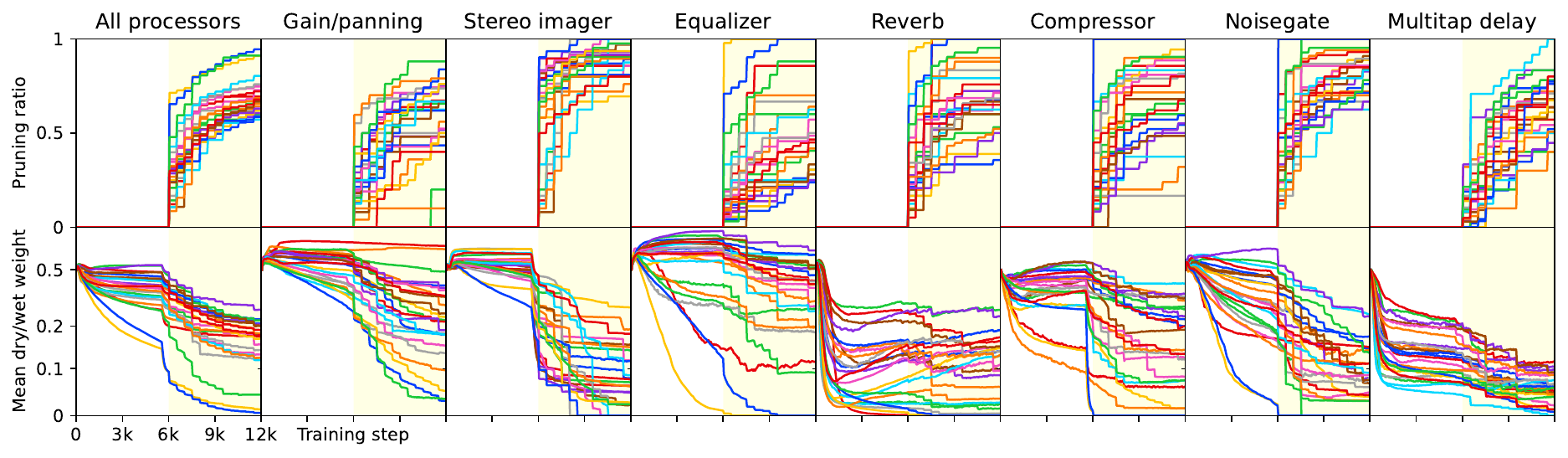}}
    \vspace{-5mm}
        \caption{
			Process of iterative pruning
			(hybrid, $\tau=0.01$). 
			  $24$ random-sampled songs are shown; 
                each color represents an individual song. 
			The upper and lower rows show the pruning ratios
			and mean dry/wet weights. 
			The yellow-shaded regions show the pruning phase.
        } 
        \label{fig:pruning-process}
    \end{center}
    \vspace{-1mm}
\end{figure*}

\subsection{Candidate Sampling}\label{section:candidate}
Implementation-wise, we have to decide how to sample a candidate set $\Vc$ 
(\ref{algo:prune:sample}, \ref{algo:prune:update-pool}),  decide when to terminate the trials (\ref{algo:prune:while}), and do some preprocessing (\ref{algo:prune:preprocess}; see colored lines in Algorithm \ref{algo:pruning}).
We explore the following $3$ approaches.

First, we can use a simple brute-force method; we can try every processor one by one: $\Vc=\{v\}$ where $v \sim V_\mathrm{pool}$. We repeat this until we try every processor in the pool $V_\mathrm{pool}$.
This granularity likely achieves high sparsity but demands a high computational cost; we should run $|V_\mathrm{pool}|$ trials for each pruning stage.

Second, to improve efficiency from the above method, we need an informed guess of each node's importance.
Intuitively, we can use each dry/wet weight $w_i$ for approximate importance as $w_i = 0$ is equivalent to pruning $v_i$.
This observation leads to the following variant, the dry/wet method, described in Algorithm \ref{algo:pruning-weight} (the colored lines highlight the correspondence with Algorithm \ref{algo:pruning}). 
For each pruning iteration, we create a set of remaining types $\Tc$ and count the processors of each type $N_t$ (\ref{algo:prune-weight:init-start}-\ref{algo:prune-weight:init-end}).
Then, for each trial, we sample a remaining type $t$ (\ref{algo:prune-weight:sample-start}) and choose the smallest-weight type-$t$ processors as candidates $\bar{V}_t$ (\ref{algo:prune-weight:sample-end}).
The proportion of chosen candidates is controlled with a per-type ratio $r_t$ initialized to $0.1$.
Then, we perform an iterative search to find the maximum proportion for each type.
When the trial succeeds, we prune those candidates and continue the search for that type $t$ (\ref{algo:prune-weight:update-success}).
If the trial fails, we perform the following: if the number of candidates is larger than $1$, we halve the ratio for the next search of the type $t$. 
Otherwise, we terminate the search for the type-$t$ nodes by removing the type from the pool $\Tc$.
This method resembles the binary search but uses a more conservative initial proportion $r_t = 0.1$ other than $0.5$. This avoids excessive pruning of early-sampled types.
Additionally, treating each type separately is motivated by the observation that the range of dry/wet values may differ for different types.

Finally, solely relying on the weight values could neglect some large-weight processors that can be pruned.
We mitigate this by combining the above two methods, running the brute-force method for every $4^\mathrm{th}$ iteration and the dry/wet method for the remaining. We call this a hybrid method, which is, along with the tolerance of $\tau = 0.01$, our default choice throughout the experiments.

\subsection{Optimization}
We keep the same objectives used in the mixing console training and add sparsity regularization $L_\mathrm{p}$, 
an $l_1$ norm of the weight $\mathbf{w}$,
to promote the pruning.
\begin{equation}
    L(\bar{\bfP}, \bfw) = \La(\bar{\bfP}, \bfw) + \alpha_\mathrm{g} \Lg(\bar{\bfP}) + \alpha_\mathrm{p} \Lp(\bfw).
\end{equation}
We begin by training the console for $6\si{k}$ steps. Following this, we perform $N_\mathrm{iter}=12$ rounds of pruning, with each round including $0.5\si{k}$ steps of fine-tuning. This results in the same total optimization steps as in the previous experiments. Although the full console optimization steps are halved, potentially leading to an increase in loss, this approach is justified due to strict resource constraints. During the initial $4\si{k}$ steps of the pruning phase, we increase the sparsity coefficient $\alpha_\mathrm{p}$ linearly from $0$ to $10^{-4}$.
\section{Results} \label{section:pruning-results}

\subsection{Sparsity and Quality Degradation} 
Same as the mixing console experiments, we report results for \texttt{MedleyDB} and \texttt{MixingSecrets} datasets separately (Table \ref{table:pruning-medley} and \ref{table:pruning-mixing-secrets}, respectively).
In the default setting, the average audio loss $\La$ is $0.301$ for the \texttt{MedleyDB} subset, which is a $0.005$ increase compared to the full consoles within the range of the tolerance threshold of $\tau = 0.01$. 
However, for the \texttt{MixingSecrets} subset, the audio loss increase is $0.018$, higher than the tolerance. 
This outcome is expected due to the shorter training time for the full console. 
The average pruning ratio, denoted as $\mu$, is $0.606$ and $0.622$, respectively, with the equalizer \texttt{e} and stereo imager \texttt{s} being the most and least retained types, respectively (e.g., $0.805$ and $0.483$ for \texttt{MedleyDB}, respectively).
Similar to the mixing console experiments, metrics other than FAD have noisy results, although the changes in values are very small (e.g., at most $0.5\si{dB}$ for SI-SDR).

\subsection{Choice of Tolerance}
Even with a very low tolerance of $\tau = 0.001$, we can nearly halve the number of processors (average pruning ratio of $\mu = 0.442$ and $0.433$ for the respective subsets). However, setting $\tau$ too high, such as $\tau = 0.1$, results in highly sparse graphs but significantly degrades the match quality. 
The default $\tau = 0.01$ has a good balance, maintaining the match while achieving reasonable sparsity.
This can be verified qualitatively with the spectrogram error plots; see the bottom 3 rows of Figure \ref{fig:spec-main} (and the latter subjective listening test results). 
There is no noticeable degradation in performance from the full consoles to the pruned graphs with $\tau = 0.001$ and $\tau = 0.01$. Also, refer to the captions in Figure \ref{fig:pruned-graphs}, which provide the pruning ratios and audio losses for the pruned graphs of the song \texttt{RockSteady} (note that the full console has an audio loss of $0.523$).

\subsection{Pruning Process} 
Figure \ref{fig:pruning-process} shows how the pruning progresses. During the first $6\si{k}$ steps of initial console optimization, the dry/wet weights adapt freely to optimize the audio loss $\La$ since there is no sparsity regularization $L_\mathrm{p}$. 
Then, as the pruning phase begins, the weight values start to decrease. 
Each graph's sparsity increases gradually while its weights adapt over time.
Note that this trend is different for different processor types, supporting the per-type candidate selection of the dry/wet and hybrid methods.

\subsection{Sampling Method Comparison} 
\begin{figure}[!t]
  \centering
  \vspace{-3mm}
  \includegraphics[width=.98\columnwidth]{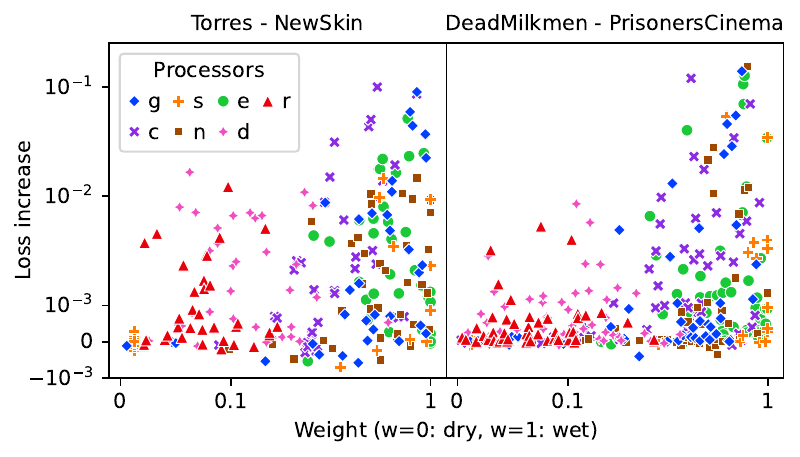}
  \vspace{-3mm}
  \caption{
	  Each node's weight and loss increase when pruned.
  }
  \label{fig:loss-vs-weight} 
\end{figure}

We compare the three proposed candidate sampling variants under the same tolerance value $\tau = 0.01$.
The brute-force method, having the highest granularity, also achieves the highest sparsity by removing $62.6\%$ and $65.4\%$ of the processors on the \texttt{MedleyDB} and \texttt{MixingSecrets} datasets, respectively. 
As fewer processors remain, it also has the worst match quality: the MRSTFT loss $\La$ increases by $0.009$ and $0.021$ from the mixing console, respectively.
The opposite results are obtained with the dry/wet method. It prunes less and degrades the quality less.
Similar results can be observed with most other validation metrics, while results with MIR feature distances are more mixed.

To investigate the differences in sparsity, we examine the relationship between each dry/wet weight $w_i$ and the loss increase caused by pruning the corresponding processor $v_i$
defined as follows,
\begin{equation}
    \Delta_i = \La(G\setminus\{v_i\}) - \La(G).
\end{equation}
Figure \ref{fig:loss-vs-weight} shows scatterplots for $2$ randomly sampled songs.
Each point $(w_i, \Delta_i)$ represents a processor after initial console training. 
Ideally, maximizing sparsity with the dry/wet method would require a monotonic relationship between $w_i$ and $\Delta_i$.
However, this is not observed, and only a weak positive correlation exists.
Also, different types are clustered in different regions, again justifying the per-type candidate selection.
Still, we cannot completely remove the weakness of the dry/wet method,
leading us to the hybrid strategy as a compromise.

\begin{figure}[t]
  \centering
  \vspace{-3mm}
  \includegraphics[width=.99\columnwidth]{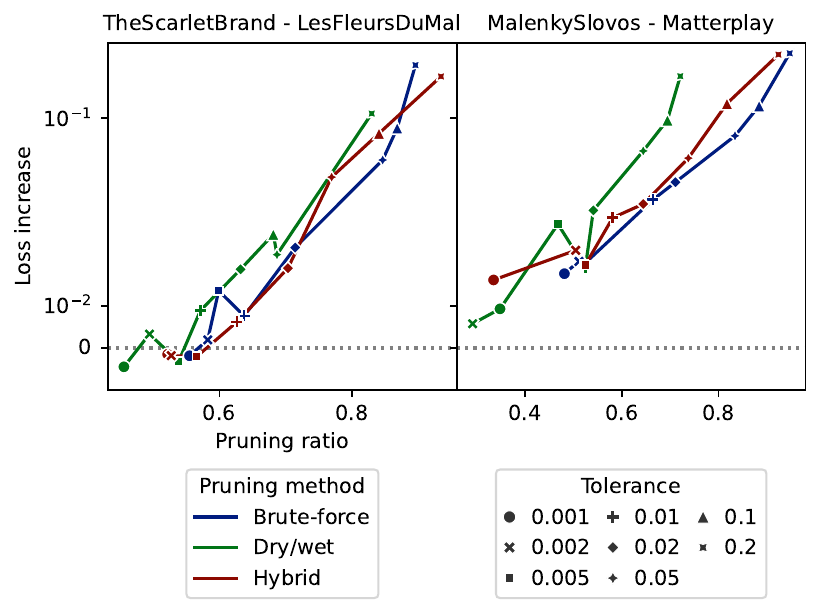}
  \vspace{-1.5mm}
  \caption{
	  	Loss increases from the mixing console and pruning ratios 
		for different pruning methods and tolerances.
  }
  \label{fig:loss-vs-ratio} 
\end{figure}

The three methods differ not only in sparsity but also in the trade-offs between sparsity and match performance. 
We can observe this by running the methods with more fine-grained tolerance values, from $0.001$ to $0.2$, and analyzing the loss-sparsity relationship.
Figure \ref{fig:loss-vs-ratio} shows the result.
The brute-force method tends to find graphs with better matches even with the same graph size (lower right region in each subplot).
However, in terms of the search time, the brute-force, dry/wet, and hybrid methods took an average of $56$, $29$, and $36$ minutes for each song, respectively, using a single RTX $3090$ GPU.
Note that this difference comes from the different number of candidate trials. 
Furthermore, it remains unclear whether this difference in the trade-off is perceptually significant, raising the question of whether the finer search is worth the cost.

\begin{figure}[t]
  \centering
  \vspace{-1mm}
  \includegraphics[width=.99\columnwidth]{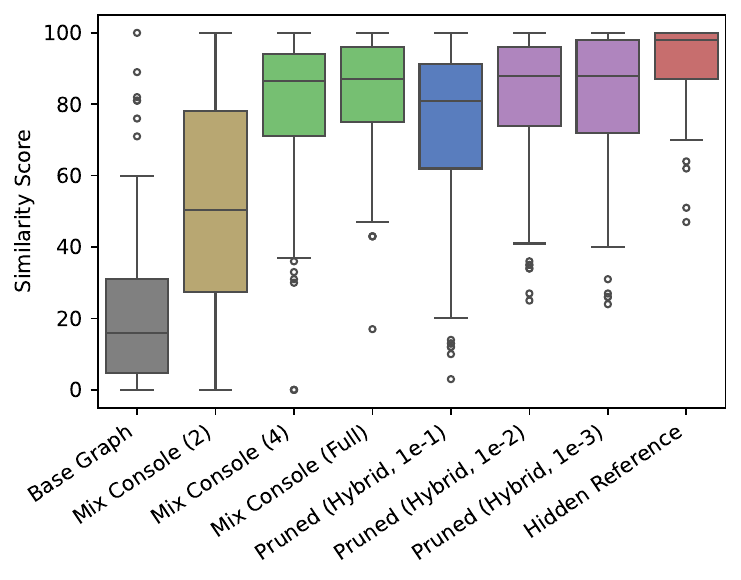}
  \vspace{-3mm}
  \caption{
	  Subjective listening test results.
  }
  \label{fig:mushra} 
\end{figure}

\subsection{Subjective Listening Test}
To further investigate the performance of our methods, we conducted a MUltiple Stimuli with Hidden Reference and Anchor (MUSHRA) test \cite{mushra}.
We asked participants to compare the original mix (reference) and the others with scores ranging from $0$ to $100$. A total of $8$ stimuli were evaluated, including (i) the hidden reference, (ii) naive summation as a low anchor, and the mixing consoles with (iii) $2$ and (iv) $4$ processor types (as in Table \ref{table:mixing-console-medley}) as mid anchors.
The remaining four are the (v) full mixing console and its pruned versions with the hybrid method and the tolerance $\tau$ of (vi) $0.1$, (vii) $0.01$, and (viii) $0.001$. 
$18$ subjects participated in the test: 
$9$ were professional mixing engineers and $6$ were amateurs ($4.5$ years of experience on average). 
The remaining $3$ had no experience in music mixing, but they are researchers in musical signal processing and familiar with the MUSHRA-style test. 
Each participant went through $2$ training and $12$ main sessions. Each session was a random $8$-second excerpt from the validation set.
We used WebMUSHRA \cite{schoeffler2018webmushra} to conduct the test remotely.
\begin{figure}[!ht]
  \centering
  \vspace{-2.5mm}
  \includegraphics[width=\columnwidth]{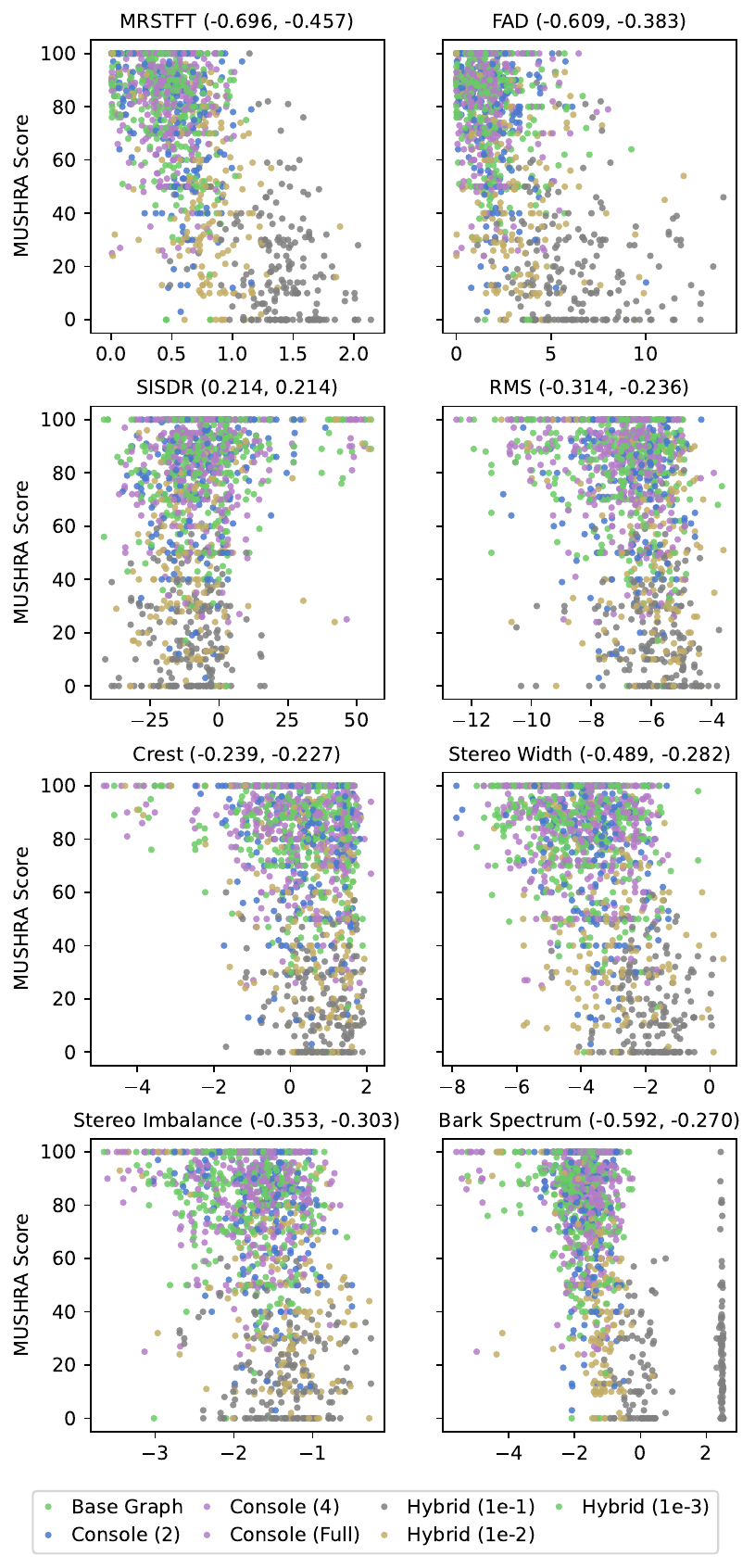}
  \vspace{-7mm}
  \caption{
	  Scatter plots of MUSHRA scores and objective metrics.
  }
  \label{fig:mushra-scatter} 
  \vspace{-2mm}
\end{figure}

Figure \ref{fig:mushra} summarizes the results. 
The hidden reference scored $92.5 \pm 1.3$ ($95\%$ confidence interval). 
As expected, the base graph received the lowest score of $21.1 \pm 2.7$. 
Note that some high-score outliers exist; this is because their corresponding excerpts were ``easy," where the naive summations sufficiently match the targets.
The mixing console with $2$ types (gain/panning and stereo imager) scored $51.2 \pm 4.0$, indicating that matching the overall loudness and stereo balance is insufficient.
By adding the equalizer and reverb, the score improved significantly to $80.5 \pm 2.6$.
The full mixing console scored $83.6 \pm 2.1$, which is the best score among tested stimuli but still significantly lower than the hidden reference ($8.9$ points).
The pruned graphs with the tolerance $\tau$ of $0.01$ and $0.001$ received almost identical scores of $82.8 \pm 2.3$ and $82.5 \pm 2.3$, respectively. Also, their score decrease from the mixing console was very small ($1.1$ maximum).
However, setting the tolerance to $0.1$ resulted in a clear degradation and scored $74.9 \pm 3.0$.
These results align well with the previous analysis: the pruning has an almost negligible effect on the perceptual similarity if the tolerance value $\tau$ is set appropriately. 
\begin{figure*}
    \begin{center}
    \captionsetup[subfloat]{captionskip=8pt}
    \subfloat[$\tau=0.001$\\$\rho = 0.45, \La=0.525$ \label{fig:pruned-0.001}]{
        \hspace{5mm}
        \includegraphics[height=.52\columnwidth]{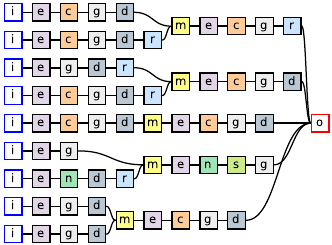}
        \hspace{5mm}
        }
    \subfloat[$\tau=0.01$\\$\rho = 0.74, \La=0.525$ \label{fig:pruned-0.01}]{
        \includegraphics[height=.52\columnwidth]{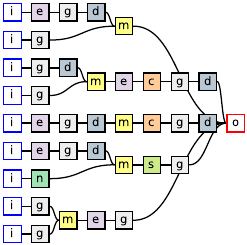}
        }
    \subfloat[$\tau=0.1$\\$\rho = 0.85, \La=0.611$
    \label{fig:pruned-0.1}]{
        \hspace{5mm}
        \includegraphics[height=.52\columnwidth]{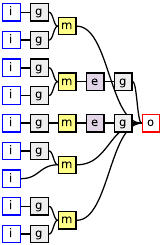} 
        \hspace{5mm}
        } 
    \caption{
		Pruning results (hybrid method) with various tolerances. 
		Song: \texttt{TablaBreakbeatScience\_RockSteady}.
	}
    \label{fig:pruned-graphs} 
    \vspace{-3mm}
    \end{center}
\end{figure*}

\begin{figure}[t]
    \begin{center}
    \vspace{-2mm}
	\setlength\tabcolsep{1pt}
	\renewcommand{\arraystretch}{1}
	\begin{tabular}{ll}
    \subfloat{
        \includegraphics[height=.395\columnwidth]{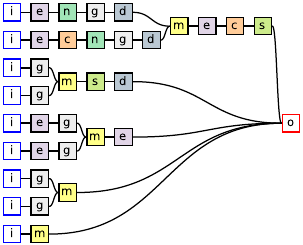}
		}
		&
    \subfloat{
        \includegraphics[height=.395\columnwidth]{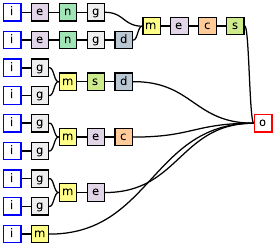}
		}
		\\
    \subfloat{
        \includegraphics[height=.395\columnwidth]{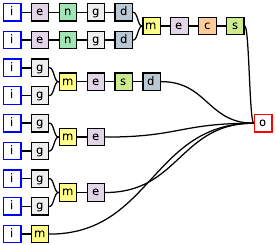}
		}
		&
    \subfloat{
        \includegraphics[height=.395\columnwidth]{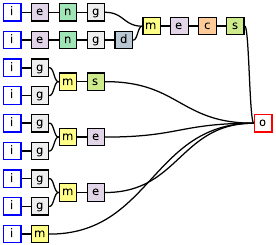}}
		\\
	\end{tabular}
    \captionsetup[subfloat]{captionskip=9pt}
	\caption{
	Graphs obtained with multiple pruning runs (default setting). 
	Song: \texttt{EthanHein\_GirlOnABridge}.    
}
    \label{fig:pruned-graphs-multiple} 
    \vspace{-2mm}
    \end{center}
\end{figure}

It could be helpful to investigate how well the loss function and evaluation metrics correlate with subjective similarity. 
Figure \ref{fig:mushra-scatter} shows scatter plots between the MUSHRA score and the loss/metrics. For each pair, we also calculated $2$ Pearson correlation coefficients (PCCs): one with the full test results and the other one that discarded the base graph samples. 
The latter was motivated by the fact that the base graph results are, in many cases, too different from the target and not of interest. 
These PCC values are also reported in Figure \ref{fig:mushra-scatter} (see the parentheses). 
The results show that the MRSTFT loss shows the strongest linear correlation, making it a suitable objective for optimization and pruning. 
The second best is the FAD score, possibly due to its ability to capture diverse features of the music mixes, even if it was not pre-trained for that purpose.
The MIR feature distances have slightly weaker correlations, which is expected as each feature reflects certain specific aspects of the mix.
The SI-SDR has the weakest correlation, possibly because of the misalignment mentioned above and the sample-level match not being perceptually important.

After the test, we asked the subjects what kind of differences they could find between the reference and the other stimuli, especially the full consoles and the pruning with $\tau \leq 0.01$.
One common response was the slight mismatch in the loudness dynamics and spatial effects.
This applied to both the full consoles and their pruned versions, indicating that the processors or optimization objectives could be further improved. 
Additionally, the participants noticed some smearing artifacts, which may have been caused by the reverbs and multitap delays while attempting to correct the misalignments.
Finally, we also conducted an informal AB test to compare the three candidate selection methods with the same tolerance $\tau = 0.01$. However, no statistically significant results were found.  
This is possible because their differences are smaller than, e.g., the differences between the hybrid method with $\tau = 0.01$ and $ \tau = 0.001$.

\subsection{Characteristics of Iterative Pruning}
\begin{figure*}
  \centering
    \subfloat[\texttt{Torres\_NewSkin} \label{fig:spec-main-full-a}]{
        \includegraphics[width=.995\columnwidth]{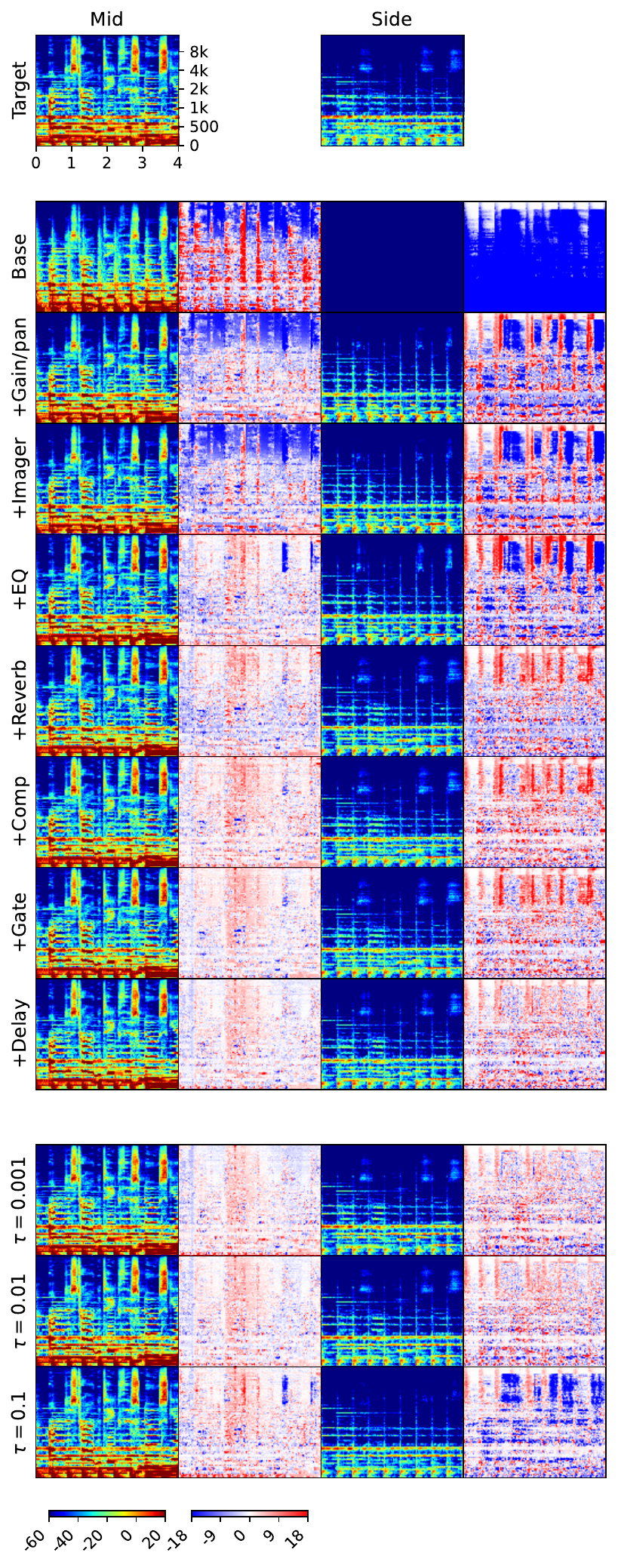}} \hspace{2mm}
    \subfloat[\texttt{TablaBreakbeatScience\_RockSteady} \label{fig:spec-main-full-b}]{
        \includegraphics[width=.995\columnwidth]{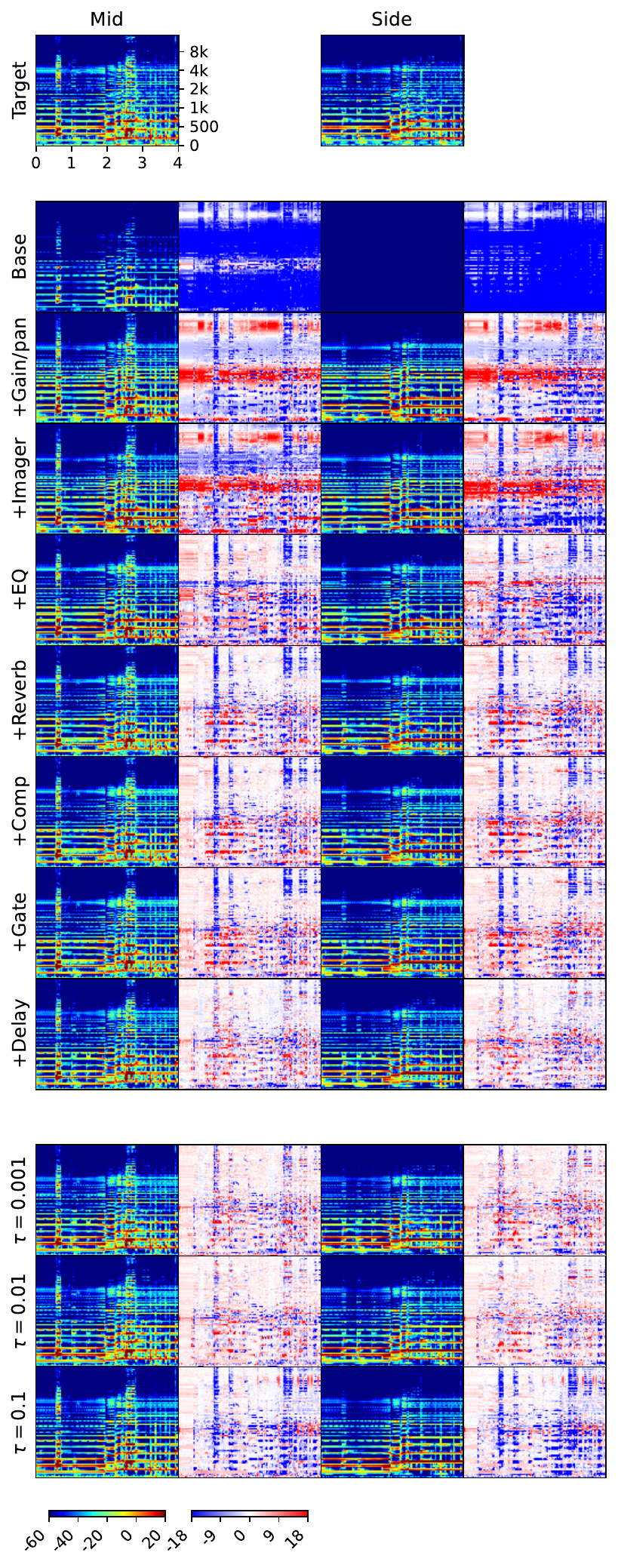}}\\
  \caption{Log-magnitude spectrograms of target mixes and matches of mixing console with different processor configurations and pruned graphs. Refer to the supplementary page for more samples. 
  }
    \label{fig:spec-main} 
\end{figure*}

Here, we provide example cases to illustrate the behavior of the pruning method.
First, we can qualitatively check that the pruning indeed removes the unnecessary processors.
Recall that adding reverbs to the song \texttt{RockSteady} did not significantly improve performance (Figure \ref{fig:spec-main-full-b}).
It is reasonable to anticipate that those reverbs will be pruned with a moderate tolerance $\tau$ accordingly.
Figure \ref{fig:pruned-graphs} shows that this is the case; only $5$ reverbs remain among $14$ when $\tau=0.001$ and zero for $\tau = 0.01$.
When $\tau=0.1$, only the gain/pannings and equalizers are left, and the others for the details are eliminated, significantly degrading the match.

It should be noted that the greedy pruning tends to yield suboptimal results.
For example, it could fail to detect redundant processors.
In Figure \ref{fig:pruned-0.01}, the bottom $2$ inputs are processed with $3$ gain/pannings; at least one of them can be removed by modifying the parameters of the remaining.
While this specific case is easy to detect and correct, it hints that sparser graphs might exist, which are harder to find.

\begin{table*}[ht]
\setlength\tabcolsep{2.7pt}
\renewcommand{\arraystretch}{.85}
\begin{center}
\caption{
Pruning results of the full dataset (default setting).
}
\small
\begin{tabular}{lcccccccccccccccc}
\toprule
& \multirow{1}[5]{*}{$L_\mathrm{a}$ $\downarrow$}
& \multicolumn{8}{c}{Pruning Ratio $\uparrow$}
& \multirow{1}[5]{*}{FAD $\downarrow$}
& \multirow{1}[5]{*}{SI-SDR $\uparrow$}
& \multicolumn{5}{c}{MIR Feature Distance $\downarrow$}
\\
\cmidrule{3-10}
\cmidrule{13-17}

& 
& $\mu$ 
& \texttt{g} 
& \texttt{s} 
& \texttt{e} 
& \texttt{r} 
& \texttt{c} 
& \texttt{n} 
& \texttt{d} 
& & & $d_\mathrm{RMS}$ & $d_\mathrm{CR}$ & $d_\mathrm{SW}$ & $d_\mathrm{SI}$ & $d_\mathrm{BS}$ 
\\

\midrule
\arrayrulecolor{lightgray}
\texttt{MedleyDB} 
& $.431$ 
& $.631$ 
& $.365$ & $.842$ & $.444$ & $.688$ & $.735$ & $.773$ & $.570$ &
$.488$ & $-4.11$ & $-5.75$ & $-.237$ & $-3.67$ & $-3.20$ & $-1.32$
\\

\texttt{MixingSecrets} 
& $.625$ & $.637$ & $.498$ & $.876$ & $.334$ & $.635$ & $.633$ & $.798$ & $.685$ &
$.621$ & $-12.2$ & $-5.03$ & $\textcolor{white}{+}.254$ & $-2.83$ & $-2.82$ & $-1.01$\\

\texttt{Internal} 
& $.434$ & $.755$ & $.705$ & $.871$ & $.550$ & $.728$ & $.854$ & $.857$ & $.723$
& $.280$ & $\textcolor{white}{+}0.28$ & $-5.88$ & $-.387$ & $-3.72$ & $-3.45$ & $-1.47$
\\

\midrule
\texttt{Total} 
& $.506$ & $.692$ & $.574$ & $.868$ & $.451$ & $.686$ & $.751$ & $.821$ & $.685$
& $.443$ & $-5.15$ & $-5.54$ & $-.120$ & $-3.37$ & $-3.17$ & $-1.27$
\\

\arrayrulecolor{black}
\bottomrule
\vspace{-5mm}
\end{tabular}
\label{table:pruning-full}
\end{center}
\end{table*}

\begin{table}[t]
\setlength\tabcolsep{2.74pt}
\renewcommand{\arraystretch}{.85}
\begin{center}
\caption{
Processor detection performance of the pruning method (default setting) with ``consistency check."
}
\small
\begin{tabular}{lcccccccc}
\toprule
& $\mu$ 
& \texttt{g} 
& \texttt{s} 
& \texttt{e} 
& \texttt{r} 
& \texttt{c} 
& \texttt{n} 
& \texttt{d} 
\\

\midrule
Accuracy $\uparrow$
& $.902$ & $.888$ & $.968$ & $.923$ & $.956$ & $.914$ & $.909$ & $.757$
\\
\arrayrulecolor{lightgray}
\midrule
Precision $\uparrow$
& $.846$ & $.848$ & $.871$ & $.954$ & $.931$ & $.848$ & $.773$ & $.682$
\\
Recall $\uparrow$
& $.900$ & $.955$ & $.903$ & $.917$ & $.931$ & $.935$ & $.874$ & $.768$\\
\midrule
F1 Score $\uparrow$
& $.872$ & $.898$ & $.887$ & $.935$ & $.931$ & $.889$ & $.820$ & $.722$\\

\arrayrulecolor{black}
\bottomrule
\end{tabular}
\label{table:consistency-check}
\vspace{-1mm}
\end{center}
\end{table}

Another notable behavior is that each pruning run of the same song results in a slightly different graph. 
We ran the pruning of \texttt{GirlOnABridge} $10$ times; 
$4$ of them are shown in Figure \ref{fig:pruned-graphs-multiple}. The resulting graphs have the number of processors ranging from $19$ to $22$ with an audio loss ranging from $0.668$ to $0.674$.
The main cause of this variance is the stochasticity of the iterative pruning; candidates sampled early are more likely to satisfy the tolerance condition and be pruned, while the remaining will adapt and be further optimized, changing their ``importance scores."

\subsection{Consistency Analysis} \label{section:consistency}
It is worth investigating how similar the pruned graph $G$ is to the ground truth. 
However, the latter is unavailable.
As a remedy, we can try a ``consistency check," where we treat the pruned result as a synthetic ground truth. Then, we can rerun the pruning with the matched mix $\hat{y}$ as a target and measure the similarity of the obtained graphs $\tilde{G}$ with the original $G$.
Because of the restrictions on the graph structure (sequential chain, fixed order, at most one processor instance per type), comparing the two graphs, $G$ and $\tilde{G}$, is equivalent to comparing the set of processors for each input and subgroup, i.e., multiple binary classification tasks.

Table \ref{table:consistency-check} summarizes the evaluation results; we report the per-type and micro-average metrics (denoted as $\mu$).
The average accuracy is $0.902$, i.e., the pruning correctly predicts about $9$ out of $10$ whether a certain source track or subgroup mix is processed with a given processor type (or not). In terms of the F1 score, the micro-average result is $0.872$. The multitap delay \texttt{d} shows the lowest result (F1 score of $0.722$), possibly due to its subtle effect. 
Note that the precision is consistently lower than recall across all types except for the equalizer \texttt{e} ($0.054$ by average). This is not so surprising; as the pruning operates with a certain tolerance, it is biased towards false negatives rather than false positives.

We note that the mean audio loss result $\La$ of this experiment was $0.112$, which is much lower than the main result ($0.422$).
This is because each target mix $\hat{y}$ is already projected to the subspace that our graph can express (shown as a teal-shaded area in Figure \ref{fig:framework-full}). 
Nevertheless, the loss value is not negligible. We suspect two reasons: the pruned graph $\tilde{G}$ is slightly different from the original $G$, and the gradient descent struggles to find better parameters.

\subsection{Full Results} \label{section:full-pruning}
\begin{figure}[t]
  \centering
  \includegraphics[width=.91\columnwidth]{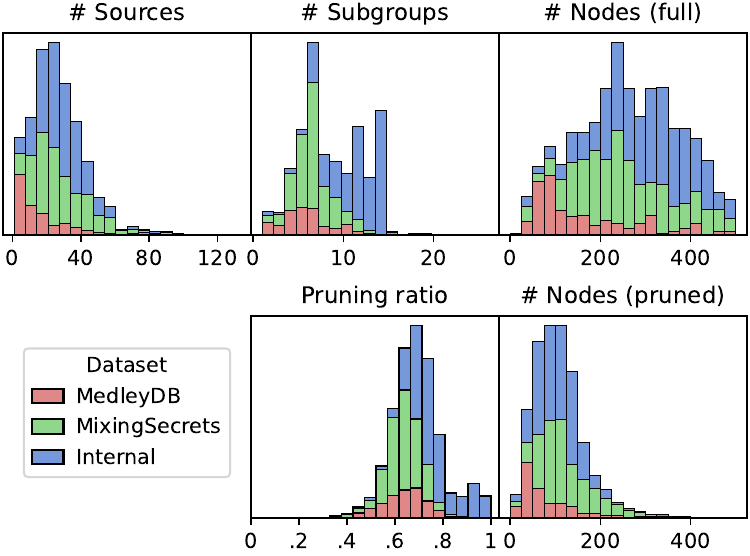}
  \caption{
	  Statistics of the consoles and pruned graphs (full data).
	Per-dataset results are stacked to form the full histograms.
  }
  \label{fig:pruned-stats} 
\end{figure}

Finally, we pruned the full dataset ensemble and collected the graph data for the downstream tasks. The evaluation results are reported in Table \ref{table:pruning-full} and they are consistent with the evaluation subset results.
Overall statistics of the songs, original mixing consoles, and their corresponding pruned graphs are shown in Figure \ref{fig:pruned-stats}.
Among the datasets, \texttt{MedleyDB} has the fewest source tracks, averaging $17.6$ per song. In contrast, the \texttt{Internal} dataset has the most ($28.8$), closely followed by \texttt{MixingSecrets} ($27.9$). 
The \texttt{Internal} dataset also tends to have more subgroups, leading to larger mixing consoles. 
We suspect these contributed to the higher sparsity of its pruned graphs; more processors were initially used to match the mixes, resulting in more redundant ones that could be pruned.
On average, $72.1$ processors ($108.5$ nodes) remained per song after pruning. Considering that each full mixing console initially contained an average of $247.6$ processors ($280.1$ nodes), this corresponds to a pruning ratio of $0.692$.

\section{Conclusion}
We formulated the reverse engineering of the music mix as a search of the audio processing graph and its processor parameters.
Then, we restricted the search to the iterative pruning of a pre-defined mixing console and derived a computationally feasible algorithm. 
Extensive evaluation results indicate that, with an appropriate tolerance, pruning rarely degrades the match quality. This was supported by the almost identical MUSHRA score of the default pruning setup to that of the full mixing console. 

On the other hand, there was a clear score difference between the hidden reference and the matches, including ones from the full mixing console.
This suggests that enhancing the console can have a significant impact on the match quality.
We chose our processors, such as the zero-phase FIR equalizer and STFT mask-based reverb, due to their simplicity and fast computation on GPUs. 
However, alternatives such as parametric equalizers \cite{nercessian2020neural} and artificial reverberation \cite{lee2022differentiable} exist.
Also, when we added the dynamic processors to the mixing consoles, the listening test results showed only a slight improvement.
Approximated ballistics could be a potential cause, as recently reported \cite{steinmetz2023high}.
We could expand the processor set to include effects like saturation \cite{colonel2022reverse} or modulation \cite{carson2023differentiable}. 
Additionally, the method did not account for time-varying parameters (e.g., automation), leading to audible mismatches, such as the inability to replicate fade-outs.

The structure of the mixing consoles could also be modified to better reflect real-world practices. 
We could add send-return loops, post-equalizers for compressors, and even multi-input or multi-output processors, such as those with sidechains or crossover filters. 
While we used already available subgroup information to construct a mixing console, it is possible to estimate it with audio feature analysis \cite{ronan2015automatic}.
We could also jointly optimize the subgrouping information.
However, we should extend the current framework, e.g., allowing learnable edge weights, to accommodate such graph connectivity.

Another critical component of the method is the optimization objective.
The analysis of the subjective listening test results showed that the current MRSTFT loss correlated most closely with the MUSHRA score among all tested metrics.
However, it has also been reported to miss certain perceptual features \cite{hayes2023review, turian2020m}. 
Exploring alternative objectives  \cite{vahidi2023mesostructures} could address these shortcomings. 

The resulting graph's sparsity and the search cost were also of interest due to the application of graph data collection. In this context, we explored three pruning variants with different granularities in search. 
The real-world results show that a more detailed search produces more accurate graphs of the match-sparsity tradeoff. However, it's not clear if this difference is significant enough to justify the extra cost, given the limited computation budget. 
We note that the final experiment, which consisted of more than a thousand individual optimization runs, took about three days, even with $10$ RTX $3090$ GPUs.

While our method finds a sparse and interpretable mixing graph, it is still unclear how closely the pruning method finds the actual underlying graph. We only have the following conservative conclusions: Our method, at least, tries to find a minimal graph that closely follows the target. 
Also, it shows reasonable reconstruction accuracy ($0.872$ F1 score) for the in-domain graphs (obtained with the same pruning method), as shown in the ``consistency check" experiment. 
Besides that, the graph's plausibility is supported by the design of mixing consoles, their processors, and several qualitative case studies.

We list possible extensions of the pruning method.
First, differentiability is not required for the pruning objective. Allowing for broader design choices, such as the MIR feature distances used in the evaluation.
Secondly, our method tends to prune processors for short-span tracks because of their limited contribution. However, their removal can sometimes be noticeable, and methods for resolving this are desirable.
Third, adopting advanced neural network pruning techniques \cite{he2023structured} could improve the candidate node selections. Finally, we can apply domain-specific methods, e.g., merging LTI processors into a single one, which could further increase sparsity. 

Other than the aforementioned extensions, we may take a radically different approach, e.g., adopting the neural architecture search by relaxing the prior restrictions on graph structures \cite{liu2018darts, ye2023fm},  reinforcement learning \cite{elton2019deep, you2018graph}. 
While these are promising, the main challenge will be balancing various aspects, such as the graph's plausibility, match quality, sparsity, and the search method's computation cost. 

This work focused on the graph search algorithm and its application in collecting graph data.
The next step will be applying this data to downstream applications. We can extend prior works on statistical analysis of music mixes \cite{wilson2015101, mourgela2024exploring, wilson2016variation} by directly assessing the mixing graphs, not just the audio features.
We can also treat the collected graphs as pseudo-labels and train neural networks to estimate the graphs (and their corresponding parameters). We can modify existing methods, e.g., automatic mixing \cite{perez2009automatic, de2013knowledge, martinez2022automatic, koszewski2023automatic, steinmetz2020diffmixconsole, steinmetz2022automix} and style transfer \cite{koo2023music}, to output the graphs, which allows the end users to interpret and control the outputs further for their needs.
While we collected graphs from an ensemble of multiple datasets, the amount of data could still be insufficient for the neural network training. Mitigating potential overfitting and improving the generalization performance are left as future work.

\section{References} 
{

\bibliographystyle{IEEEtran}
\bibliography{refs.bib}

\begin{thebibliography}{10}
\providecommand{\url}[1]{#1}
\csname url@samestyle\endcsname
\providecommand{\newblock}{\relax}
\providecommand{\bibinfo}[2]{#2}
\providecommand{\BIBentrySTDinterwordspacing}{\spaceskip=0pt\relax}
\providecommand{\BIBentryALTinterwordstretchfactor}{4}
\providecommand{\BIBentryALTinterwordspacing}{\spaceskip=\fontdimen2\font plus
\BIBentryALTinterwordstretchfactor\fontdimen3\font minus \fontdimen4\font\relax}
\providecommand{\BIBforeignlanguage}[2]{{%
\expandafter\ifx\csname l@#1\endcsname\relax
\typeout{** WARNING: IEEEtran.bst: No hyphenation pattern has been}%
\typeout{** loaded for the language `#1'. Using the pattern for}%
\typeout{** the default language instead.}%
\else
\language=\csname l@#1\endcsname
\fi
#2}}
\providecommand{\BIBdecl}{\relax}
\BIBdecl

\bibitem{pestana2014intelligent}
P.~D. Pestana and J.~D. Reiss, ``Intelligent audio production strategies informed by best practices,'' in \emph{Audio Engineering Society 53rd International Conference on Semantic Audio}, 2014.

\bibitem{everardo2017towards}
F.~Everardo, ``Towards an automated multitrack mixing tool using answer set programming,'' in \emph{14th Sound and Music Computing Conference}, 2017.

\bibitem{ronan2017analysis}
D.~Ronan, H.~Gunes, and J.~D. Reiss, ``Analysis of the subgrouping practices of professional mix engineers,'' in \emph{Audio Engineering Society Convention}, vol. 142, 2017.

\bibitem{bittner2014medleydb}
R.~M. Bittner, J.~Salamon, M.~Tierney, M.~Mauch, C.~Cannam, and J.~P. Bello, ``{MedleyDB}: A multitrack dataset for annotation-intensive mir research.'' in \emph{Proceedings of the 24th International Society for Music Information Retrieval (ISMIR) Conference}, vol.~14, 2014, pp. 155--160.

\bibitem{bittner2016medleydb}
R.~M. Bittner, J.~Wilkins, H.~Yip, and J.~P. Bello, ``{MedleyDB} 2.0: New data and a system for sustainable data collection,'' in \emph{Proceedings of the 24th International Society for Music Information Retrieval (ISMIR) Conference, Late Breaking Demo}, 2016, p.~36.

\bibitem{senior2018mixing}
M.~Senior, \emph{Mixing Secrets for the Small Studio}.\hskip 1em plus 0.5em minus 0.4em\relax Routledge, 2018.

\bibitem{wilson2015101}
A.~Wilson and B.~Fazenda, ``101 mixes: A statistical analysis of mix-variation in a dataset of multi-track music mixes,'' in \emph{Audio Engineering Society Convention 139}.\hskip 1em plus 0.5em minus 0.4em\relax Audio Engineering Society, 2015.

\bibitem{mourgela2024exploring}
A.~Mourgela, E.~Quinton, S.~Bissas, J.~D. Reiss, and D.~Ronan, ``Exploring trends in audio mixes and masters: Insights from a dataset analysis,'' \emph{arXiv:2412.03373}, 2024.

\bibitem{wilson2016variation}
A.~Wilson and B.~Fazenda, ``Variation in multitrack mixes: analysis of low-level audio signal features,'' \emph{Journal of the Audio Engineering Society (JAES)}, vol.~64, no. 7/8, pp. 466--473, 2016.

\bibitem{de2017ten}
B.~De~Man, J.~Reiss, and R.~Stables, ``Ten years of automatic mixing,'' in \emph{3rd Workshop on Intelligent Music Production}, 2017.

\bibitem{perez2009automatic}
E.~Perez-Gonzalez and J.~Reiss, ``Automatic gain and fader control for live mixing,'' in \emph{IEEE Workshop on Applications of Signal Processing to Audio and Acoustics}, 2009, pp. 1--4.

\bibitem{de2013knowledge}
B.~De~Man and J.~D. Reiss, ``A knowledge-engineered autonomous mixing system,'' in \emph{Audio Engineering Society Convention 135}, 2013.

\bibitem{steinmetz2020diffmixconsole}
C.~J. Steinmetz, J.~Pons, S.~Pascual, and J.~Serrà, ``Automatic multitrack mixing with a differentiable mixing console of neural audio effects,'' in \emph{IEEE International Conference on Acoustics, Speech, and Signal Processing (ICASSP)}, 2021, pp. 71--75.

\bibitem{lee2023blind}
S.~Lee, J.~Park, S.~Paik, and K.~Lee, ``Blind estimation of audio processing graph,'' in \emph{IEEE International Conference on Acoustics, Speech, and Signal Processing (ICASSP)}, 2023, pp. 1--5.

\bibitem{barchiesi2010reverse}
D.~Barchiesi and J.~Reiss, ``Reverse engineering of a mix,'' \emph{Journal of the Audio Engineering Society (JAES)}, vol.~58, no. 7/8, pp. 563--576, 2010.

\bibitem{colonel2021reverse}
J.~T. Colonel and J.~Reiss, ``Reverse engineering of a recording mix with differentiable digital signal processing,'' \emph{The Journal of the Acoustical Society of America}, vol. 150, no.~1, pp. 608--619, 2021.

\bibitem{colone2023reverse}
------, ``Reverse engineering a nonlinear mix of a multitrack recording,'' \emph{Journal of the Audio Engineering Society (JAES)}, vol.~71, no.~9, pp. 586--595, 2023.

\bibitem{engel2020ddsp}
J.~Engel, L.~H. Hantrakul, C.~Gu, and A.~Roberts, ``{DDSP}: differentiable digital signal processing,'' in \emph{International Conference on Learning Representations (ICLR)}, 2020.

\bibitem{hayes2023review}
B.~Hayes, J.~Shier, G.~Fazekas, A.~McPherson, and C.~Saitis, ``A review of differentiable digital signal processing for music \& speech synthesis,'' \emph{Frontiers in Signal Process.}, p. 1284100, 2023.

\bibitem{hybridconsole}
``The mixing console --- split, inline and hybrids,'' \href{https://steemit.com/sound/@jamesub/the-mixing-console-split-inline-and-hybrids}{https://steemit.com/sound/@jamesub/the-mixing-console-split-inline-and-hybrids}, accessed: 2024-02-26.

\bibitem{paszke2019pytorch}
A.~Paszke, S.~Gross, F.~Massa \emph{et~al.}, ``{PyTorch}: An imperative style, high-performance deep learning library,'' \emph{Conference on Neural Information Processing Systems (NeurIPS)}, 2019.

\bibitem{castellano1997iterative}
G.~Castellano, A.~M. Fanelli, and M.~Pelillo, ``An iterative pruning algorithm for feedforward neural networks,'' \emph{IEEE transactions on Neural networks}, vol.~8, no.~3, pp. 519--531, 1997.

\bibitem{liu2018darts}
H.~Liu, K.~Simonyan, and Y.~Yang, ``{DARTS}: Differentiable architecture search,'' in \emph{International Conference on Learning Representations (ICLR)}, 2019.

\bibitem{ye2023fm}
Z.~Ye, W.~Xue, X.~Tan, Q.~Liu, and Y.~Guo, ``Nas-fm: Neural architecture search for tunable and interpretable sound synthesis based on frequency modulation,'' in \emph{International Joint Conference on Artificial Intelligence (IJCAI)}.\hskip 1em plus 0.5em minus 0.4em\relax International Joint Conferences on Artificial Intelligence Organization, 2023, p. 5869–5877.

\bibitem{uzrad2024diffmoog}
N.~Uzrad, O.~Barkan, A.~Elharar \emph{et~al.}, ``Diffmoog: a differentiable modular synthesizer for sound matching,'' \emph{arXiv:2401.12570}, 2024.

\bibitem{lee2024searching}
S.~Lee, M.~Mart{\'\i}nez-Ram{\'\i}rez, W.-H. Liao \emph{et~al.}, ``Searching for music mixing graphs: a pruning approach,'' in \emph{Proceedings of the International Conference on Digital Audio Effects (DAFx)}, 2024.

\bibitem{lee2024grafx}
------, ``{GRAFX}: an open-source library for audio processing graphs in {P}y{T}orch,'' in \emph{Proceedings of the International Conference on Digital Audio Effects (DAFx), Demo}, 2024.

\bibitem{kilgour2018fr}
K.~Kilgour, M.~Zuluaga, D.~Roblek, and M.~Sharifi, ``Fr{\`e}chet audio distance: A metric for evaluating music enhancement algorithms,'' \emph{arXiv:1812.08466}, 2018.

\bibitem{le2019sdr}
J.~Le~Roux, S.~Wisdom, H.~Erdogan, and J.~R. Hershey, ``Sdr--half-baked or well done?'' in \emph{IEEE International Conference on Acoustics, Speech, and Signal Processing (ICASSP)}.\hskip 1em plus 0.5em minus 0.4em\relax IEEE, 2019, pp. 626--630.

\bibitem{vanka2024diff}
S.~Vanka, C.~Steinmetz, J.-B. Rolland, J.~Reiss, and G.~Fazekas, ``Diff-mst: Differentiable mixing style transfer,'' in \emph{Proceedings of the International Society for Music Information Retrieval Conference (ISMIR)}.\hskip 1em plus 0.5em minus 0.4em\relax San Francisco, USA: Int. Society for Music Information Retrieval (ISMIR), 2024.

\bibitem{man2014analysis}
B.~Man, B.~Leonard, R.~King, and J.~D. Reiss, ``An analysis and evaluation of audio features for multitrack music mixtures,'' in \emph{Proceedings of International Society for Music Information Retrieval Conference (ISMIR)}, 2014.

\bibitem{mushra}
B.~Series, ``Method for the subjective assessment of intermediate quality level of audio systems,'' International Telecommunication Union Radio Communication Assembly, Geneva, CH, Standard, Mar. 2014.

\bibitem{schoeffler2018webmushra}
M.~Schoeffler, S.~Bartoschek, F.-R. Stöter, and p, ``web{MUSHRA} — a comprehensive framework for web-based listening tests. journal of open research software,'' \emph{Journal of Open Research Software}, vol.~6, p.~8, 2018.

\bibitem{caspe2022ddx7}
F.~Caspe, A.~McPherson, and M.~Sandler, ``{DDX7}: Differentiable {FM} synthesis of musical instrument sounds,'' in \emph{Proceedings of the International Society for Music Information Retrieval Conference (ISMIR)}, 2022.

\bibitem{Mitcheltree_2021}
C.~Mitcheltree and H.~Koike, ``{SerumRNN}: Step by step audio {VST} effect programming,'' in \emph{Artificial Intelligence in Music, Sound, Art and Design}, 2021, pp. 218--234.

\bibitem{guo2023automatic}
J.~Guo and B.~McFee, ``Automatic recognition of cascaded guitar effects,'' in \emph{Proceedings of the International Conference on Digital Audio Effects (DAFx)}, 2023.

\bibitem{take2024audio}
O.~Take, K.~Watanabe, T.~Nakatsuka \emph{et~al.}, ``Audio effect chain estimation and dry signal recovery from multi-effect-processed musical signals,'' in \emph{Proceedings of the International Conference on Digital Audio Effects (DAFx)}, 2024.

\bibitem{ramirez2021differentiable}
M.~A. Mart{\'\i}nez-Ram{\'\i}rez, O.~Wang, P.~Smaragdis, and N.~J. Bryan, ``Differentiable signal processing with black-box audio effects,'' in \emph{IEEE International Conference on Acoustics, Speech, and Signal Processing (ICASSP)}, 2021.

\bibitem{steinmetz2022style}
C.~J. Steinmetz, N.~J. Bryan, and J.~D. Reiss, ``Style transfer of audio effects with differentiable signal processing,'' \emph{Journal of the Audio Engineering Society (JAES)}, vol.~70, no.~9, pp. 708--721, 2022.

\bibitem{colonel2022reverse}
J.~T. Colonel, M.~Comunit{\`a}, and J.~Reiss, ``Reverse engineering memoryless distortion effects with differentiable waveshapers,'' in \emph{Audio Engineering Society Convention 153}, 2022.

\bibitem{carson2023differentiable}
A.~Carson, S.~King, C.~V. Botinhao, and S.~Bilbao, ``Differentiable grey-box modelling of phaser effects using frame-based spectral processing,'' in \emph{Proceedings of the International Conference on Digital Audio Effects (DAFx)}, 2023.

\bibitem{nercessian2020neural}
S.~Nercessian, ``Neural parametric equalizer matching using differentiable biquads,'' in \emph{Proceedings of the International Conference on Digital Audio Effects (DAFx)}, 2020, pp. 265--272.

\bibitem{lee2022differentiable}
S.~Lee, H.-S. Choi, and K.~Lee, ``Differentiable artificial reverberation,'' \emph{IEEE/ACM Transactions on Audio, Speech and Language Processing (TASLP)}, vol.~30, pp. 2541--2556, 2022.

\bibitem{he2023structured}
Y.~He and L.~Xiao, ``Structured pruning for deep convolutional neural networks: A survey,'' \emph{IEEE Transactions on Pattern Analysis and Machine Intelligence (TPAMI)}, vol.~46, no.~5, p. 2900–2919, 2024.

\bibitem{elsken2019neural}
T.~Elsken, J.~H. Metzen, and F.~Hutter, ``Neural architecture search: A survey,'' \emph{Journal of Machine Learning Research (JMLR)}, vol.~20, no.~55, pp. 1--21, 2019.

\bibitem{elton2019deep}
D.~C. Elton, Z.~Boukouvalas, M.~D. Fuge, and P.~W. Chung, ``Deep learning for molecular design—a review of the state of the art,'' \emph{Molecular Systems Design \& Engineering}, vol.~4, no.~4, pp. 828--849, 2019.

\bibitem{you2018graph}
J.~You, B.~Liu, Z.~Ying, V.~Pande, and J.~Leskovec, ``Graph convolutional policy network for goal-directed molecular graph generation,'' \emph{Conference on Neural Information Processing Systems (NeurIPS)}, 2018.

\bibitem{fu2023flashfftconv}
D.~Y. Fu, H.~Kumbong, E.~Nguyen, and C.~R{\'e}, ``{FlashFFTConv}: Efficient convolutions for long sequences with tensor cores,'' \emph{International Conference on Learning Representations (ICLR)}, 2023.

\bibitem{giannoulis2012digital}
D.~Giannoulis, M.~Massberg, and J.~D. Reiss, ``Digital dynamic range compressor design—a tutorial and analysis,'' \emph{Journal of the Audio Engineering Society (JAES)}, vol.~60, no.~6, pp. 399--408, 2012.

\bibitem{hayes2023sinusoidal}
B.~Hayes, C.~Saitis, and G.~Fazekas, ``Sinusoidal frequency estimation by gradient descent,'' in \emph{IEEE International Conference on Acoustics, Speech, and Signal Processing (ICASSP)}, 2023, pp. 1--5.

\bibitem{bengio2013estimating}
Y.~Bengio, N.~L{\'e}onard, and A.~Courville, ``Estimating or propagating gradients through stochastic neurons for conditional computation,'' \emph{arXiv:1308.3432}, 2013.

\bibitem{yamamoto2020parallel}
R.~Yamamoto, E.~Song, and J.-M. Kim, ``Parallel wavegan: A fast waveform generation model based on generative adversarial networks with multi-resolution spectrogram,'' in \emph{IEEE International Conference on Acoustics, Speech, and Signal Processing (ICASSP)}, 2020, pp. 6199--6203.

\bibitem{wright2020perceptual}
A.~Wright and V.~V{\"a}lim{\"a}ki, ``Perceptual loss function for neural modeling of audio systems,'' in \emph{IEEE International Conference on Acoustics, Speech, and Signal Processing (ICASSP)}, 2020, pp. 251--255.

\bibitem{steinmetz2020auraloss}
C.~J. Steinmetz and J.~D. Reiss, ``auraloss: {A}udio focused loss functions in {PyTorch},'' in \emph{Digital Music Research Network One-day Workshop (DMRN+15)}, 2020.

\bibitem{loshchilov2017decoupled}
I.~Loshchilov and F.~Hutter, ``Decoupled weight decay regularization,'' \emph{International Conference on Learning Representations (ICLR)}, 2019.

\bibitem{steinmetz2023high}
C.~J. Steinmetz, T.~Walther, and J.~D. Reiss, ``High-fidelity noise reduction with differentiable signal processing,'' in \emph{Audio Engineering Society Convention 155}, 2023.

\bibitem{ronan2015automatic}
D.~Ronan, H.~Gunes, D.~Moffat, and J.~D. Reiss, ``Automatic subgrouping of multitrack audio,'' in \emph{Proceedings of the International Conference on Digital Audio Effects (DAFx)}, 2015.

\bibitem{turian2020m}
J.~Turian and M.~Henry, ``I’m sorry for your loss: Spectrally-based audio distances are bad at pitch,'' in \emph{``I Can't Believe It's Not Better!'' Conference on Neural Information Processing Systems (NeurIPS) workshop}, 2020.

\bibitem{vahidi2023mesostructures}
C.~Vahidi, H.~Han, C.~Wang, M.~Lagrange, G.~Fazekas, and V.~Lostanlen, ``Mesostructures: Beyond spectrogram loss in differentiable time--frequency analysis,'' \emph{Journal of the Audio Engineering Society (JAES)}, 2023.

\bibitem{martinez2022automatic}
M.~A. Mart{\'\i}nez-Ram{\'\i}rez, W.-H. Liao, G.~Fabbro, S.~Uhlich, C.~Nagashima, and Y.~Mitsufuji, ``Automatic music mixing with deep learning and out-of-domain data,'' in \emph{Proceedings of the International Society for Music Information Retrieval Conference (ISMIR)}, 2022.

\bibitem{koszewski2023automatic}
D.~Koszewski, T.~G{\"o}rne, G.~Korvel, and B.~Kostek, ``Automatic music signal mixing system based on one-dimensional wave-u-net autoencoders,'' \emph{EURASIP Journal on Audio, Speech, and Music Processing (JASM)}, vol. 2023, no.~1, pp. 1--17, 2023.

\bibitem{steinmetz2022automix}
C.~J. Steinmetz, S.~S. Vanka, M.~A. Mart{\'\i}nez-Ram{\'\i}rez, and G.~Bromham, \emph{Deep Learning for Automatic Mixing}.\hskip 1em plus 0.5em minus 0.4em\relax ISMIR, Dec. 2022.

\bibitem{koo2023music}
J.~Koo, M.~A. Mart{\'\i}nez-Ram{\'\i}rez, W.-H. Liao \emph{et~al.}, ``Music mixing style transfer: A contrastive learning approach to disentangle audio effects,'' in \emph{IEEE International Conference on Acoustics, Speech, and Signal Processing (ICASSP)}.\hskip 1em plus 0.5em minus 0.4em\relax IEEE, 2023, pp. 1--5.

\end{thebibliography}
}
\end{document}